\documentclass[pra,superscriptaddress,showpacs,longbibliography,reprint]{revtex4-2}

\usepackage[colorlinks=true,linkcolor=blue,urlcolor=blue,citecolor=blue,pdfusetitle]{hyperref}
\usepackage{color}
\usepackage[utf8]{inputenc}
\usepackage[usenames,dvipsnames]{xcolor}
\usepackage{amsmath}
\usepackage{amssymb}
\usepackage{graphicx}
\graphicspath{{graphics/}}
\usepackage{epsfig}
\usepackage{dcolumn}
\usepackage{bm}
\usepackage{mathrsfs}
\usepackage{multirow}
\usepackage[all]{xy}
\usepackage{pbox}
\usepackage{lipsum}
\usepackage{verbatim}
\usepackage{braket}
\usepackage{dsfont}
\usepackage{array}
\usepackage{makecell}
\usepackage{tabularx}
\usepackage{isotope}
\usepackage{xr}
\usepackage[normalem]{ulem}

\usepackage[colorlinks=true,linkcolor=blue,urlcolor=blue,citecolor=blue,pdfusetitle]{hyperref}

\begin{document}

\title{Parameter estimation with one- and two-time measurements on the emission field of the boundary time crystal}
\author{Albert Cabot}
\affiliation{Institute for Cross-Disciplinary Physics and Complex Systems (IFISC) UIB-CSIC, Campus Universitat Illes Balears, 07122, Palma de Mallorca, Spain.}
\affiliation{Institut für Theoretische Physik and Center for Integrated Quantum Science and Technology, Universität Tübingen, Auf der Morgenstelle 14, 72076 Tübingen, Germany.}
\author{Federico Carollo}
\affiliation{Dipartimento di Fisica, Sapienza Università di Roma, Piazzale Aldo Moro 2, 00185 Rome, Italy.}
\affiliation{Centre for Fluid and Complex Systems, Coventry University, Coventry, CV1 2TT, UK.}
\author{Igor Lesanovsky}
\affiliation{Institut für Theoretische Physik and Center for Integrated Quantum Science and Technology, Universität Tübingen, Auf der Morgenstelle 14, 72076 Tübingen, Germany.}
\affiliation{School of Physics and Astronomy, University of Nottingham, Nottingham, NG7 2RD, UK.}
\affiliation{Centre for the Mathematics and Theoretical Physics of Quantum Non-Equilibrium Systems,
University of Nottingham, Nottingham, NG7 2RD, UK.}

\begin{abstract}
Many-body quantum systems can exhibit collective effects that enhance the sensitivity of parameter estimation protocols. An example is provided by resonantly driven two-level atoms subject to collective dissipation, which can display a transition between a stationary phase and a time-crystal one. Previous work has shown that the light emitted in the time-crystal  phase can be harnessed for parameter estimation using continuous monitoring protocols, such as photon counting or homodyne detection,  which under ideal conditions yield a quadratic enhancement of sensitivity with the number of particles. In this work, we explore what is the minimal information about the emission field that needs to be accessed in order to resolve collective effects and exploit them for parameter estimation.  We show that, for short probing times, a single-time measurement of the emission field already captures the collective behavior emerging at the nonequilibrium transition. In contrast, within the time‑crystal phase, exploiting collective effects requires at least two-time measurements.  To this end, we introduce a family of correlated intensity measurements that extract the relevant information and can be implemented using an interferometric setup. While the ultimate sensitivity bound remains size independent, as recently established within the framework of noisy quantum metrology, our analysis shows that these protocols utilize collective effects to yield a transient increase in sensitivity with particle number.
\end{abstract}

\maketitle

\section{Introduction}

Quantum correlations and collective phenomena in many-body systems can serve as a resource for sensing applications \cite{Degen2017,Braun2018}. Examples include protocols to detect small displacements or weak electromagnetic fields in systems of trapped ions \cite{Gilmore2021}, or Rydberg atoms \cite{Facon2016,Jing2020}. 
In such settings, the number of particles $N$ acts as a metrological resource: quantum correlations can improve the scaling of the estimation variance from the standard quantum limit $N^{-1}$ to the Heisenberg limit $N^{-2}$.
This enhancement has been observed, for instance, in protocols exploiting so-called N00N states for phase estimation \cite{Giovannetti2004}, although their experimental preparation and robustness against decoherence remain challenging \cite{Degen2017}.
Spin‑squeezed states offer another route to surpass the standard quantum limit \cite{Wineland1992,Kitagawa1993,Ma2011}, and can be generated in large atomic ensembles coupled to optical cavities through cavity-mediated interactions or measurement‑based feedback protocols \cite{Ma2011,Barberena2024,Leroux2010,Hosten2016,Cox2016,Braverman2019}.
These examples illustrate that dissipative and noisy effects can actually play an active role in the design of sensing protocols. 

Sensing via continuous monitoring \cite{Wiseman2009} exploits the emission signal of a dissipative quantum system to perform parameter estimation  \cite{Catana2012,Gammelmark2013,Cortez2017,Gross2018,Shankar2019,Rossi2020,Angelatos2021,Fallani2022,Ilias2022,Rinaldi2024,Ilias2024,Radaelli2024arxiv}. This approach allows one to gather the information the system emits through  different decay channels. Fundamental sensitivity bounds for continuous-monitoring schemes can be computed  based on the Quantum Fisher Information (QFI) and have been derived    \cite{Gammelmark2014,Guta2015,Catana2015,Macieszczak2016,Albarelli2017} together with general optimal measurement strategies \cite{Yang2023,Godley2023}. The combination of driving, dissipation and interactions in many-body systems can give rise to  emergent nonequilibrium phases and phase transitions \cite{Kessler2012,Lee2013b,Minganti2018}, whose enhanced susceptibilities can be exploited for sensing, as shown both theoretically \cite{Macieszczak2016,Fernandez2017,Ilias2022,Gandia2023,Pavlov2023,Montenegro2023,Ilias2024,Alushi2025} and experimentally \cite{Ding2022}. Nonequilibrium phases, such as synchronized phases of atomic dipoles, can further increase the coherence time, which is a key resource for phase  estimation \cite{Xu2014,Xu2015,Shankar2017}. In quantum optical platforms, this collective behavior leaves clear signatures in the emitted light which can be subsequently used for sensing applications.

\begin{figure}[t!]
 \centering
 \includegraphics[width=1\columnwidth]{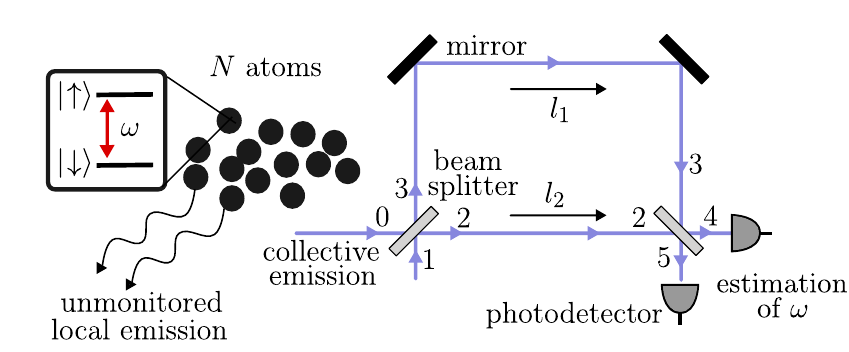}
 \caption{{\bf Sketch of the system and measurement protocol.} We consider an ensemble of $N$ two-level atoms, emitting collectively and driven at resonance with Rabi frequency $\omega$. We are interested in the estimation of small variations of $\omega$ by analyzing the light that is emitted collectively.  The two-time measurement protocol could be implemented through a Mach-Zehnder interferometer. The collective emission of the system is input into arm '0', while vacuum is input into arm '1'. Photodetectors, are placed at the output arms '4' and '5'. The difference in path length $l_1-l_2$ is chosen such that we can probe the field emitted at  two times of interest.}
 \label{fig_cartoon}
\end{figure}

Cooperative resonance‑fluorescence models constitute a paradigmatic example in which collective phenomena governs the properties of the emitted light \cite{Agarwal1977,Drummond1978,Carmichael1980}. This kind of models have been recently revisited in the context of time crystals \cite{Iemini2018} and nonequilibrium superradiant transitions observed in dense atomic clouds \cite{Ferioli2023,Agarwal2024,Goncalves2024}. Time-crystal phases are characterized by spontaneous time-translation symmetry breaking in the thermodynamic limit, which can occur in a variety of settings and due to very different mechanisms \cite{Else2020,Gong2018,Iemini2018,Wang2018,Tucker2018,Gambetta2019,Buca2019,Buca2019b,Zhu2019,Kessler2019,Riera2020,Lazarides2020,Lledo2020,Seibold2020,Buonaiuto2021,Hajdusek2022,Krishna2023,Cabot2024b}. Experimental observations include driven-dissipative atomic clouds and condensates \cite{Kessler2021,Liu2025}, in which signatures of their collective phenomena are imprinted in the emitted light. Time crystals have also been  previously studied in the context of sensing \cite{Montenegro2023,Cabot2024,Iemini2024,Gribben2024}. For example,  Ref. \cite{Cabot2024} showed that the emission field of the boundary time crystal (BTC) \cite{Iemini2018} can exhibit Heisenberg scaling of the QFI with particle number when estimating the driving strength. Ref. \cite{Mattes2025} later proved that ideal photoncounting and homodyne detection suffice to saturate this ultimate QFI, while Ref. \cite{o2025quantum}, using tools from noisy quantum metrology \cite{Sekatski2017,demkowicz2017adaptative,zhou2018achieving,wan2022bounds}, demonstrated that detection inefficiencies destroy this asymptotic scaling.

In this work, we consider the boundary time crystal and its emission field with the goal of identifying the minimal portions of the output field that must be probed in order to access the different collective effects displayed by the system, and of assessing the parameter estimation capabilities of protocols restricted to such minimal samples. We address these questions by focusing on measurement strategies that rely solely on information contained in the emission field at a single time or at two times at most. In order to do so in a general fashion, we consider an explicit description of the emission field in terms of the discrete time input-output formalism (see Sec. \ref{Sec_discrete}), and compute the QFI of different parts of the emission field. By analyzing these measurements across the phase diagram,  we identify  minimal measurement schemes that allow one to exploit the collective phenomena displayed by the system, and we compare their sensitivity with the ultimate limit provided by noisy quantum metrology (Sect. \ref{Sec_bounds}). 
In particular, we show that measuring the field at a single time already suffices to harness the nonequilibrium transition from the stationary to the time‑crystal phase as a sensing resource. In contrast, the oscillatory dynamics within the time‑crystal phase is only a resource for sensing when measurements probe at least two‑time correlations of the emission field. As we show, such sensitivity can be accessed by measuring the intensity at the output of an interferometer, as  depicted in Fig. \ref{fig_cartoon} and explained in detail in Sect. \ref{Sec_measurements}.  Finally, in Sec. \ref{Sec_localdecay}, we address the effects of local decay channels on this measurement scheme.

\section{The model}\label{Sec_Model}

The system we consider consists of  an ensemble of $N$ two-level atoms undergoing collective processes as described by the following Markovian master equation for the state of the system $\rho$ ($\hbar=1$ hereafter):
\begin{equation}\label{ME}
\partial_t{\rho}=\mathcal{L}\rho=-i\omega[{S}_\mathrm{x},{\rho}]+\Gamma \left({S}_-{\rho}{S}_+-\frac{1}{2}\{ {S}_+{S}_-,{\rho}\} \right).    
\end{equation}
Here, $\mathcal{L}$ is the Liouvillian superoperator and we have defined the total angular momentum operators   ${S}_\alpha=\frac{1}{2}\sum_{j=1}^N {\sigma}_\alpha^{(j)}$ ($\alpha=$x,y,z) with  ${\sigma}^{(j)}_\alpha$ being the Pauli matrices associated with atom $j$ and ${S}_\pm={S}_\mathrm{x}\pm i{S}_\mathrm{y}$. An extension of Eq. (\ref{ME}) to include the effects of local spontaneous emissions from the atoms is  introduced and analyzed in Sec. \ref{Sec_localdecay}.

The  model in Eq. (\ref{ME}), also known as the boundary time crystal \cite{Iemini2018},  provides a tractable scenario where  collective nonequilibrium phenomena emerge and can be resolved through the statistics of the emitted light \cite{Carmichael1980} or through continuous monitoring protocols \cite{Link2019,Cabot2023}. 
For large system sizes, the system displays a crossover between two dynamical regimes separated around $\omega_\mathrm{c}=N\Gamma/2$ \cite{Carmichael1980}. For $\omega<\omega_\mathrm{c}$, the system displays a fast overdamped relaxation to the stationary state. For $\omega>\omega_\mathrm{c}$, the system  displays an oscillatory decay to the stationary state. The quality factor of these oscillations increases linearly with system size \cite{Carmichael1980,Buonaiuto2021}. In the thermodynamic limit, these oscillations become nondecaying resulting in the emergence of a time-crystal phase \cite{Iemini2018}, which is described by the mean-field equations of motion \cite{Carollo2022}. In turn, the crossover becomes a sharp nonequilibrium phase transition. We note that in order to properly analyze the thermodynamic  limit one should rescale the decay rate with system size, i.e. $\Gamma\to \Gamma/N$ \cite{Iemini2018,Carollo2022}. In this work we focus on finite systems for which the above rescaling is not necessary. This also allows us to directly connect with experimental systems based on atom-cavity setups \cite{Norcia2016b,Muniz2020}  or dense pencil-shaped atomic clouds \cite{Ferioli2023}.

\section{Discrete time representation of the emission field}\label{Sec_discrete}

\begin{figure}[t!]
 \centering
 \includegraphics[width=0.7\columnwidth]{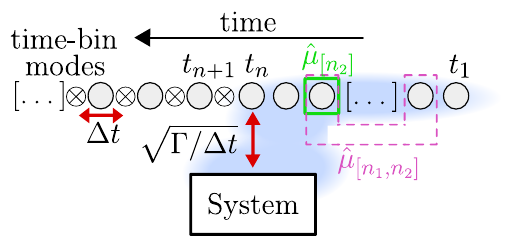}
 \caption{{\bf Sketch of the discrete time representation of the input-output field.} The input-output field is discretized in time bins of length $\Delta t$ such that $t_n=n\Delta t$ with $n=1,2,\dots$ Each bin is represented by an independent bosonic mode (time-bin modes), that interacts with the system for a time window $\Delta t$ and with strength $\sqrt{\Gamma/\Delta t}$. In the sketch the system has already interacted with the first $n$ time-bin modes. Hence, the output field is formed by the time-bin modes $[1,n]$, which are generally in a nonseparable state which is also correlated with the system itself (here pictorially represented as a blue shadow). The input field is given by bins $[n+1,\infty)$, which are in a product vacuum state. The reduced state of the time-bin mode $n_2$, which has already interacted with the system, is denoted by $\hat{\mu}_{[n_2]}$. The reduced joint state of time-bin modes $n_1$ and $n_2$ ($n_2>n_1$), that have already interacted with the system, is denoted by $\hat{\mu}_{[n_1,n_2]}$. In this work we focus on parameter estimation based on the information contained in these reduced states of the emitted light field.}
 \label{fig_timebins}
\end{figure}

In this section we introduce a discrete time description of the system-emission dynamics coarse grained over the time scales in which the master equation in Eq. (\ref{ME}) is valid. This approach, based on the input-output formalism  \cite{Gardiner2004}, provides a convenient way to analyze the information transferred from the system to the output field \cite{Gammelmark2014,Gross2018,Regidor2021,Ciccarello2022,Yang2023}. An illustration of the idea behind this formalism is presented in Fig. \ref{fig_timebins}. The input-output field is discretized in time bins of length $\Delta t$, the so-called {\it time-bin modes}, each of them corresponding to an independent bosonic mode. When measuring all of them we recover the case of ideal continuous monitoring, while when we trace them out we recover the master equation dynamics. In this work, we are interested in the reduced states of just one and two time-bin modes $\hat{\mu}_{[n_2]}$ and $\hat{\mu}_{[n_1,n_2]}$ [cf. Fig. \ref{fig_timebins}], which contain the output of the system at one or two times, respectively. These allow us, for instance, to analyze the photo detection statistics at these times, using e.g. an interferometer (see Fig. \ref{fig_cartoon}). The reader who is not interested in the mathematical details of this description can jump to Sec. \ref{Sec_bounds}, in which we discuss the fundamental bounds to sensing using the whole emitted light or only the reduced light field states $\hat{\mu}_{[n_2]}$ and $\hat{\mu}_{[n_1,n_2]}$.

\subsection{System and environment model} 

We employ the so-called quantum  input-output formalism \cite{Gardiner2004}, in which  the Hamiltonian in the laboratory frame for the system and environment is given by the sum of the following terms:
\begin{equation}
\begin{split}
&H_\mathrm{S}(t)=\omega_0 S_\mathrm{z}+\frac{\omega}{2} \big( S_+ e^{-i\omega_0 t}+S_- e^{+i\omega_0 t}\big),\\ &H_\mathrm{E}=\int d\nu\, \nu a^\dagger(\nu)a(\nu),\\
&V_\mathrm{SE}=i\sqrt{\frac{\Gamma}{2\pi}}\int_{\omega_0-\mathcal{B}}^{\omega_0+\mathcal{B}}d\nu [S_- a^\dagger(\nu)-S_+a(\nu)].
\end{split}
\end{equation}
Here $a(\nu)$, $a^\dagger(\nu)$ are bosonic annihilation and creation operators, $[a(\nu),a^\dagger(\nu')]=\delta(\nu-\nu')$, representing the light field at frequency $\nu$. 2$\mathcal{B}$ is the bandwidth of the system environment coupling, which is small compared to $\omega_0$ ($\omega_0\gg \omega,\Gamma$ is also assumed) \cite{Gross2018}. In the interaction picture with respect to $\omega_0 S_\mathrm{z}+H_\mathrm{E}$, we obtain the following  Hamiltonian:
\begin{equation}
H_\mathrm{SE}(t)=\omega S_\mathrm{x}+i\sqrt{\Gamma}[S_- a^\dagger(t)-S_+a(t)],     
\end{equation}
where
\begin{equation}
a(t)=\frac{1}{\sqrt{2\pi}}\int_{\omega_0-\mathcal{B}}^{\omega_0+\mathcal{B}}d\nu\, a(\nu) e^{-i(\nu-\omega_0)t}.     
\end{equation}

We consider the  dynamics  coarse grained on a timescale much larger than $\mathcal{B}^{-1}$. At this scale, the field operators  $a(t)$  satisfy $[a(t),a^\dagger(t')]=\delta(t-t')$ \cite{Gardiner2004}. A discrete-time dynamics can be derived by considering discrete time steps that are small compared to the relaxation timescales of Eq. (\ref{ME}), but much larger than the coarse-graining timescale $\Delta t \gg \mathcal{B}^{-1}$ \cite{Gross2018,Regidor2021,Yang2023,Ciccarello2022}. Then the time window $[0,T)$ is split in $M$ time bins, denoted with the label $[n]=[t_{n-1},t_n)$, where $t_n=n\Delta t$ and $T=M\Delta t$. Following Refs. \cite{Gross2018,Ciccarello2022}, for each time bin we define the coarse-grained field operators:
\begin{equation}
 b_{[n]}=\frac{1}{\sqrt{\Delta t}}\int_{t_{n-1}}^{t_n} d\tau a(\tau),    
\end{equation}
which satisfy  bosonic commutation relations $[b_{[n]}, b^\dagger_{[n']}]=\delta_{nn'}$. Each of these bosonic modes defines an independent piece of the light field, or {\it time-bin mode}, for which there is an associatied  Fock space: $|\sigma_n\rangle= b^{\dagger\sigma}_{[n]}|0_n\rangle/\sqrt{\sigma!}$ with $\sigma_n=0,1,2,\dots$ The time evolution can then be approximated (up to order $\Delta t$) by:
\begin{equation}\label{unitary_n}
\begin{split}
U_\mathrm{SE}^{[n]}(\Delta t)\approx&1-i\omega S_\mathrm{x}\Delta t+\sqrt{\Gamma \Delta t}(S_- b_{[n]}^\dagger-\text{H.c.})\\
&-\frac{\Gamma\Delta t}{2}(S_+S_- b_{[n]} b_{[n]}^\dagger +\text{H.c.})\\
&+\frac{\Gamma \Delta t}{2}(S_-^2b^{\dagger2}_{[n]}+\text{H.c.}).    
\end{split}
\end{equation}
Considering the initial uncorrelated system-field state  $|\Psi(0)\rangle=|\psi_\mathrm{S}(0)\rangle\otimes\left(\otimes_{n=1}^\infty |0_n\rangle\right)$, we recover the master equation (\ref{ME}) after evolving each time step with the unitary operator in Eq. (\ref{unitary_n}), tracing out the time-bin modes, and performing the continuum limit or small-$\Delta t$ regime ($\Delta t\to0$) \cite{Ciccarello2022,Yang2023}. When instead explicitly considering the emission field, we can restrict the time-bin modes Hilbert space  to $\sigma_n=0,1$ as long as $\Delta t$ is small enough \cite{Gross2018,Regidor2021,Yang2023,Ciccarello2022}.  This is analogous to the assumption made in photocounting unravellings in which a time step $\Delta t$ is chosen small enough such that there is at maximum one detection per time bin. Following the discrete time-evolution implemented by Eq. (\ref{unitary_n}), the joint system-emission  state at time $T$ can be written as:
\begin{equation}\label{SE_MPS}
|\Psi(T)\rangle =\sum_{\{\sigma_n\}} K^{\sigma_M}_{[M]}\dots K^{\sigma_2}_{[2]}K^{\sigma_1}_{[1]}|\psi_\mathrm{S}(0)\rangle\otimes|\sigma_1\sigma_2\dots\sigma_M\rangle,
\end{equation}
where $\{\sigma_n\}$ denote all possible combinations of $\sigma_n=0,1$. In this expression we have used the Kraus operators
\begin{equation}\label{Kraus_def}
K_{[n]}^{\sigma_n}=\langle \sigma_n|U^{[n]}_\mathrm{SE}(\Delta t)|0_n\rangle,   
\end{equation}
which, to leading order in $\Delta t$, read:
\begin{equation}\label{Kraus_ME}
\begin{split}
&K_0\equiv K^0_{[n]}\approx 1-i\omega S_\mathrm{x}\Delta t-\frac{\Gamma}{2}S_+S_-\Delta t,\\
&K_1\equiv K^1_{[n]}\approx \sqrt{\Delta t\Gamma}\,S_-,
\end{split}
\end{equation}
and thus do not depend on $n$. Note that quantum trajectories corresponding to ideal photocounting or homodyne detection can be obtained by performing the corresponding measurements on the time-bin modes appearing in the state in Eq. (\ref{SE_MPS}), see e.g. \cite{Gross2018,Yang2023}.

\subsection{Reduced state for  time-bin modes} 

We are interested in the information that is contained in small portions of the output light field, as given by only few time-bin modes. When tracing out time-bin modes, the dynamics given by  Eq. (\ref{unitary_n}) can be conveniently implemented in terms of Kraus operators.  Defining $\hat{\varrho}(T)=|\Psi(T)\rangle\langle \Psi(T)|$, the corresponding reduced state of the system  $\rho(T)=\text{Tr}_\mathrm{E}\{\hat{\varrho}(T)\}$ can also be obtained by successive applications of the CPTP map:
\begin{equation}\label{CM_0}
\rho(n\Delta t)=\mathcal{E}\rho([n-1]\Delta t), \quad \mathcal{E}\rho=K_0\rho K_0^\dagger+K_1\rho K_1^\dagger,    
\end{equation}
to the initial condition, which in the small-$\Delta t$ regime converges to the dynamics of Eq. (\ref{ME}).  This  approach allows us to efficiently study an intermediate situation in which instead of carrying on the full system-emission state $\hat{\varrho}(T)$, we just keep track of the system and few time-bin modes. In the simplest case, we can trace out all time-bin modes but the $n_1$-th one, obtaining the following state ($T=M\Delta t$):
\begin{equation}
\hat{\varrho}_{[n_1]}(T)=\text{Tr}_{\mathrm{E}\setminus [n_1]}\{ \hat{\varrho}(T) \}.   
\end{equation}
Similarly, we can define the reduced state of system and two time-bin modes:
\begin{equation}
\hat{\varrho}_{[n_1,n_2]}(T)=\text{Tr}_{\mathrm{E}\setminus [n_1,n_2]}\{ \hat{\varrho}(T) \}.
\end{equation}
The  dynamics of these joint states can be efficiently simulated by including the corresponding time-bin modes degrees of freedom (e.g. modes $n_1$ and $n_2$) in the system Hamiltonian (see Appendix \ref{Appendix_collision_model}). In turn, if we are interested in the information contained only in the time-bin modes, we can further trace out the system obtaining the reduced states for one and two time-bin modes:
\begin{equation}
\begin{split}
\hat{\mu}_{[n_1]}(T)&=\text{Tr}_\mathrm{S}\{ \hat{\varrho}_{[n_1]}(T)\}, \\
\hat{\mu}_{[n_1,n_2]}(T)&=\text{Tr}_\mathrm{S}\{ \hat{\varrho}_{[n_1,n_2]}(T) \}.  
\end{split}
\end{equation}

In the small-$\Delta t$ regime, we can use the short time expansion of the interaction unitary, i.e. Eq. (\ref{unitary_n}), to obtain approximate expressions for the reduced states of the time-bin modes (see Appendix \ref{Appendix_collision_model}). For the case of one time-bin mode that has interacted at time $t_1=n_1\Delta t$ with the system, we obtain:
\begin{equation}\label{linear_state}
\begin{split}
  \hat{\mu}_{[n_1]}\approx& \,\,\big(1-\Gamma\Delta t \langle S_+S_-\rangle_{t_1}\big)|0_{n_1}\rangle\langle 0_{n_1}|\\
&+ \sqrt{\Gamma\Delta t}\big( \langle S_-\rangle_{t_1} |1_{n_1}\rangle \langle 0_{n_1}|
+\mathrm{H.c.}\big)\\
&+ \Gamma\Delta t \langle S_+S_-\rangle_{t_1}|1_{n_1}\rangle\langle 1_{n_1}| +\mathcal{O}(\Delta t^{3/2}),
\end{split}
\end{equation}
where $\mathcal{O}(\Delta t^{3/2})$ denotes  terms of  order $\Delta t^{3/2}$. The notation $\langle \dots \rangle_t$ indicates that the expectation  values are computed over the system reduced state at time $t$. From Eq. (\ref{linear_state}) we observe that  $\Gamma\Delta t$ is paired with the expectation value of a two-body operator $\langle S_+S_-\rangle_{t}$, which can potentially scale as $N^2$. This means that the $\Delta t$ for which the short-time expansion is valid can vary with system size.

In the case of two time-bin modes, the first interacting with the system at time $t_1$ and the second at time $t_2=n_2\Delta t$, we obtain:
\begin{equation}\label{analytical_two_ancilla}
\begin{split}
\hat{\mu}_{[n_1,n_2]}\approx&\,\ \hat{\mu}_{[n_1]}\otimes|0_{n_2}\rangle\langle0_{n_2}|+|0_{n_1}\rangle\langle0_{n_1}|\otimes\hat{\mu}_{[n_2]}\\
&+\Gamma\Delta t \langle S_-(\tau) S_-\rangle_{t_1}|1_{n_1}1_{n_2}\rangle\langle 0_{n_1}0_{n_2}|\\
&+\Gamma\Delta t \langle S_+ S_+(\tau)\rangle_{t_1}|0_{n_1}0_{n_2}\rangle\langle 1_{n_1}1_{n_2}|\\
&+\Gamma\Delta t \langle S_+(\tau) S_-\rangle_{t_1}|1_{n_1}0_{n_2}\rangle\langle 0_{n_1}1_{n_2}|\\
&+\Gamma\Delta t \langle S_+ S_-(\tau)\rangle_{t_1}|0_{n_1}1_{n_2}\rangle\langle 1_{n_1}0_{n_2}|\\
&-|0_{n_1}0_{n_2}\rangle\langle 0_{n_1}0_{n_2}| +\mathcal{O}(\Delta t^{3/2}),
\end{split}
\end{equation}
where we have defined $\tau=(n_2-n_1-1)\Delta t$, and we have used  the definition of two-time correlations \cite{CarmichaelBook}:
\begin{equation}
\begin{split}
\langle O_1(\tau)O_2\rangle_{t_1}=\text{Tr}\{O_1 e^{\mathcal{L}\tau} \big[O_2\rho(t_1) \big] \},\\
\langle O_1 O_2(\tau)\rangle_{t_1}=\text{Tr}\{O_2 e^{\mathcal{L}\tau} \big[\rho(t_1) O_1\big] \}.
\end{split}
\end{equation}

In the small-$\Delta t$ regime, the reduced state of one time-bin mode contains information of the system at just a single time, while if we keep more time-bin modes we have access to system multi-time correlations [see Fig. \ref{fig_cartoon}]. In fact, in this limit, the reduced state of the system converges to the one described by the master equation (\ref{ME}) \cite{Ciccarello2022}, while observables computed on one or two time-bin modes converge to quantities depending only on one- or two-time correlations of the system as computed with Eq. (\ref{ME}) (see Appendix \ref{Appendix_collision_model}). Therefore, in this limit, the time-bin modes describe pieces of the output light field of the system modeled by the master equation (\ref{ME})  \cite{Ciccarello2022}.

\section{Sensitivity bounds for parameter estimation}\label{Sec_bounds}

\subsection{Parameter estimation with the full emission field}

Performing measurements on the system-emission joint state [Eq. (\ref{SE_MPS})] we can implement a sensing protocol to estimate a parameter of interest. The precision at which a parameter can be estimated through any protocol is fundamentally bounded by the quantum Fisher information (QFI) of this joint state through the quantum Cramér-Rao bound \cite{Gammelmark2014,Catana2012,Guta2015}. We denote by $\Delta g(T)$ the variance on the estimated value of the parameter $g$ over a measurement time window $T$. When the protocol makes use of an unbiased estimator, this can be expressed as:
\begin{equation}
\Delta g(T)\geq \frac{1}{{\mathcal{F}_\mathrm{SE}(g,T)}}
\end{equation}
where $\mathcal{F}_\mathrm{SE}(g,T)$ is the QFI of the system-emission joint state at time $T$ and parameter $g$. 
When the Liouvillian is gapped, the long-time behavior of this QFI is linear in $T$ \cite{Gammelmark2014,Catana2012,Macieszczak2016}:
\begin{equation}
\lim_{T\to\infty} \frac{\mathcal{F}_\mathrm{SE}(g,T)}{T}= \mathcal{F}_\mathrm{SE}(g).
\end{equation}
Moreover, for long times the main contribution to this QFI comes from the information encoded in the output light field \cite{Yang2023}, as this grows with the measurement time $T$. 
For the BTC, the most interesting results are displayed in the time-crystal phase in which the QFI displays the  many-body enhanced scaling  $\mathcal{F}_\mathrm{SE}(\omega,T)\propto TN^2$ \cite{Cabot2024}. This can be saturated with ideal photon counting and homodyne continuous monitoring alone \cite{Mattes2025}, although this scaling is lost once the environment is not fully monitored (see below) \cite{o2025quantum}. In the following we investigate how collective effects manifest in the QFI of the reduced states of one and two time-bin modes.

\begin{figure}[t!]
 \centering
 \includegraphics[width=1\columnwidth]{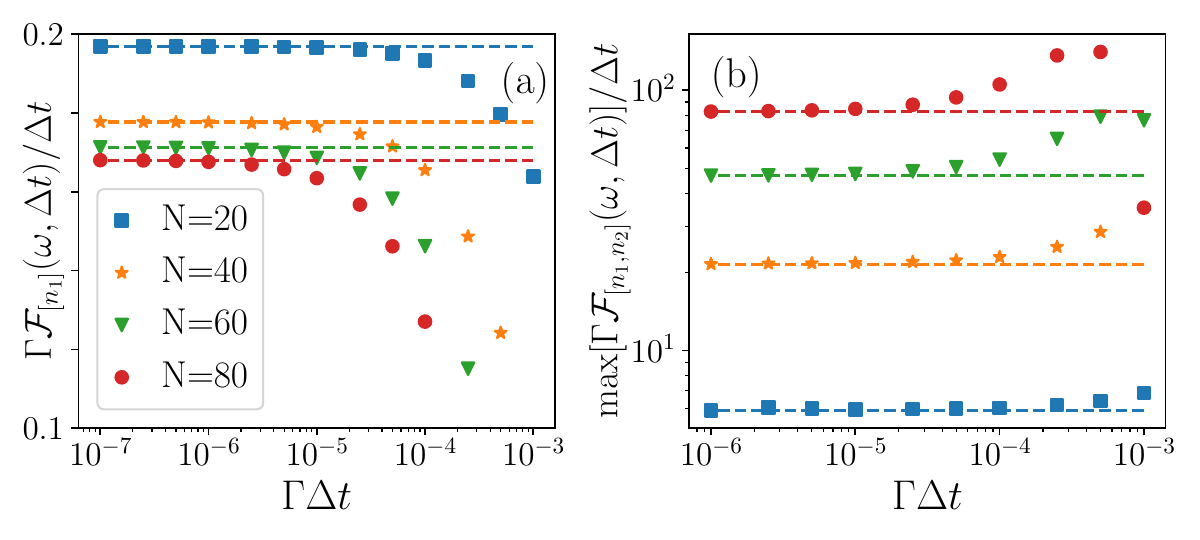}
 \caption{{\bf QFI in the small-$\Delta t$ regime.} (a) $\mathcal{F}_{[n_1]}(\omega,\Delta t)/\Delta t$ for the one time-bin reduced state (in the stationary state, $n_1\gg1$). (b) $\mathcal{F}_{[n_1,n_2]}(\omega,\Delta t)/\Delta t$ for the two time-bin reduced state optimized over the time difference between time bins. In both panels $\omega=2\omega_\mathrm{c}$. The dashed horizontal lines are a guide to the eye. The results of this figure indicate that for sufficiently small $\Delta t$ the QFI scales linearly with the interaction time $\Delta t$.}
 \label{fig_QFI_convergence}
\end{figure}

\subsection{Parameter estimation with one and two time-bin modes}

The reduced state of time-bin modes is generally a mixed state. For the one time-bin mode reduced state, the QFI is given by (see e.g. \cite{Braun2018}):
\begin{equation}
\mathcal{F}_{[n_1]}(g,\Delta t)=8 \lim_{\delta g \to 0} \frac{1-\mathbb{F}(\hat{\mu}_{[n_1]}|_{g-\delta g},\hat{\mu}_{[n_1]}|_{g+\delta g})}{(2\delta g)^2},   
\end{equation}
where $\hat{\mu}_{[n_1]}|_{g\pm\delta g}$  denote the one time-bin mode state obtained  evolving the full dynamics  for the parameter values $g\pm\delta g$, respectively. This formula makes use of the Fidelity $\mathbb{F}(\hat{\mu}_1,\hat{\mu}_2)=\text{Tr}\big[\sqrt{\sqrt{\hat{\mu}_1}\hat{\mu}_2\sqrt{\hat{\mu}_1}}\big]$, which quantifies how sensitive is the state to a small parameter change. Similarly, in the two time-bin mode case, the QFI is given by:
\begin{equation}
\mathcal{F}_{[n_1,n_2]}(g,\Delta t)=8 \lim_{\delta g \to 0} \frac{1-\mathbb{F}(\hat{\mu}_{[n_1,n_2]}|_{g-\delta g},\hat{\mu}_{[n_1,n_2]}|_{g+\delta g})}{(2\delta g)^2}, 
\end{equation}
where  $\hat{\mu}_{[n_1,n_2]}|_{g\pm\delta g}$ correspond to the two time-bin modes reduced states obtained evolving the dynamics with parameter values $g \pm\delta g$. 

In the small-$\Delta t$ regime, we find these QFI to display a linear scaling with $\Delta t$:
\begin{equation}\label{QFI_timebins_scaling}
\begin{split}
\lim_{\Delta t\to 0} \frac{\mathcal{F}_{[n_1]}(g,\Delta t)}{\Delta t}&= \mathcal{F}_{[n_1]}(g),\\
\lim_{\Delta t\to 0} \frac{\mathcal{F}_{[n_1,n_2]}(g,\Delta t)}{\Delta t}&= \mathcal{F}_{[n_1,n_2]}(g). 
\end{split}
\end{equation}
We show this numerically in Fig. \ref{fig_QFI_convergence} for estimating the parameter $\omega$, and considering $n_1$ large enough such that the system is in the stationary state $\rho_\mathrm{ss}$. In Fig. \ref{fig_QFI_convergence} (a) we illustrate the one time-bin mode case, while in (b) the two time-bin mode case. In the latter, we show the QFI optimized over the time of the second bin, $n_2$. 
Similarly to other quantities, we observe that the larger is $N$ the smaller $\Delta t$ needs to be  in order to find convergence to the short time behavior.

In the following we focus on the estimation of the parameter $\omega$, and we study $\mathcal{F}_{[n_1]}(\omega,\Delta t)$ and $\mathcal{F}_{[n_1,n_2]}(\omega,\Delta t)$ along the phase diagram, varying $N$ and for values of $\Delta t$ such that the short time expansion leading to Eqs. (\ref{linear_state}) and (\ref{analytical_two_ancilla}) is valid. In turn, we compare the short-time QFI per unit of time $\mathcal{F}_{[n_1]}(\omega)$, $\mathcal{F}_{[n_1,n_2]}(\omega)/2$, with the QFI rate of the full system-emission joint state  $\mathcal{F}_\mathrm{SE}(\omega)$. Our goal is to assess  whether the collective effects enhancing  full system-emission QFI manifest when probing the field at just one time or two times.

\subsection{QFI for single-time measurements of the emission field}

\begin{figure}[t!]
 \centering
 \includegraphics[width=1\columnwidth]{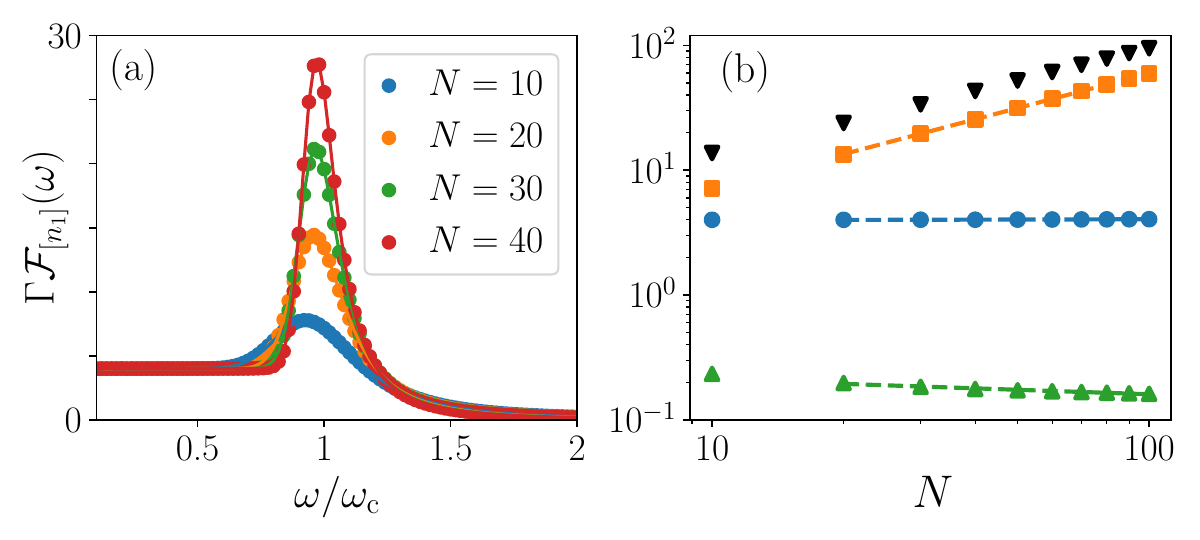}
 \caption{{\bf QFI per unit of time for one time-bin mode.} (a) $\mathcal{F}_{[n_1]}(\omega)$ varying $\omega/\omega_\mathrm{c}$ and $N$ in the long-time limit, $n_1\gg 1$. (b) Scaling of $\mathcal{F}_{[n_1]}(\omega)$ with $N$ for $\omega/\omega_\mathrm{c}=0.5$ (blue circles), $\omega/\omega_\mathrm{c}=1$ (orange squares) and $\omega/\omega_\mathrm{c}=2$ (green triangles). The dashed lines correspond to a fit $N^\alpha$ of the largest $N$ points, with exponents $\alpha=(0.01,0.93,-0.12)$ for $\omega/\omega_\mathrm{c}=(0.5,1,2)$, respectively. Black triangles correspond to  $\mathcal{F}_{\mathrm{SE}}(\omega)$ for $\omega/\omega_\mathrm{c}=1$ obtained in Ref. \cite{Cabot2024}. 
In this figure we have fixed $\Gamma\Delta t=10^{-5}$.}
 \label{fig_QFI_one_ancilla}
\end{figure}

We begin analyzing the case of a single time-bin mode reduced state for $n_1\gg1$, such that we probe the long-time statistics. In Fig. \ref{fig_QFI_one_ancilla} (a) we show the QFI per unit of time in the small-$\Delta t$ regime, $\mathcal{F}_{[n_1]}(\omega)$, varying $\omega/\omega_\mathrm{c}$ and for different system sizes $N$. Well into the overdamped regime, $\omega/\omega_\mathrm{c}<1$, we observe the QFI to be constant and independent of $N$. In this case $\mathcal{F}_{[n_1]}(\omega)$ coincides with $\mathcal{F}_{\mathrm{SE}}(\omega)$. This is because the system and emission field are in a product state, and the emission statistics is Poissonian \cite{Cabot2024}. The resulting value of the QFI per unit of time is $\mathcal{F}_{[n_1]}(\omega)\approx 4/\Gamma$ \cite{Cabot2024}. The QFI displays a peak at the phase transition, while it displays the smallest values in the time-crystal phase. In this sense, the QFI for probing the emission field at one time is qualitatively similar to the QFI of the stationary state of the  system (see Ref. \cite{Montenegro2023}).

At the phase transition point we observe the QFI to display a scaling $\propto N^{0.93}$ in the small-$\Delta t$ regime [see Fig. \ref{fig_QFI_one_ancilla} (b)]. We also compare $\mathcal{F}_{[n_1]}(\omega)$ with $\mathcal{F}_{\mathrm{SE}}(\omega)$. We observe that  by just measuring one time-bin mode a similar behavior with $N$ is obtained. This suggests that the collective effects enhancing the QFI of the full system-emission joint state are already imprinted on a single time-bin mode. In contrast, in the oscillatory regime, Fig. \ref{fig_QFI_one_ancilla} points out that measuring the emission field at just one time  and when the system is in the stationary state is not useful. The advantage of the time-crystal phase comes from the dynamical correlations, which is not captured by just one time-bin mode in the long-time limit.

\subsection{QFI for  two-time measurements of the emission field}

\begin{figure}[t!]
 \centering
 \includegraphics[width=1\columnwidth]{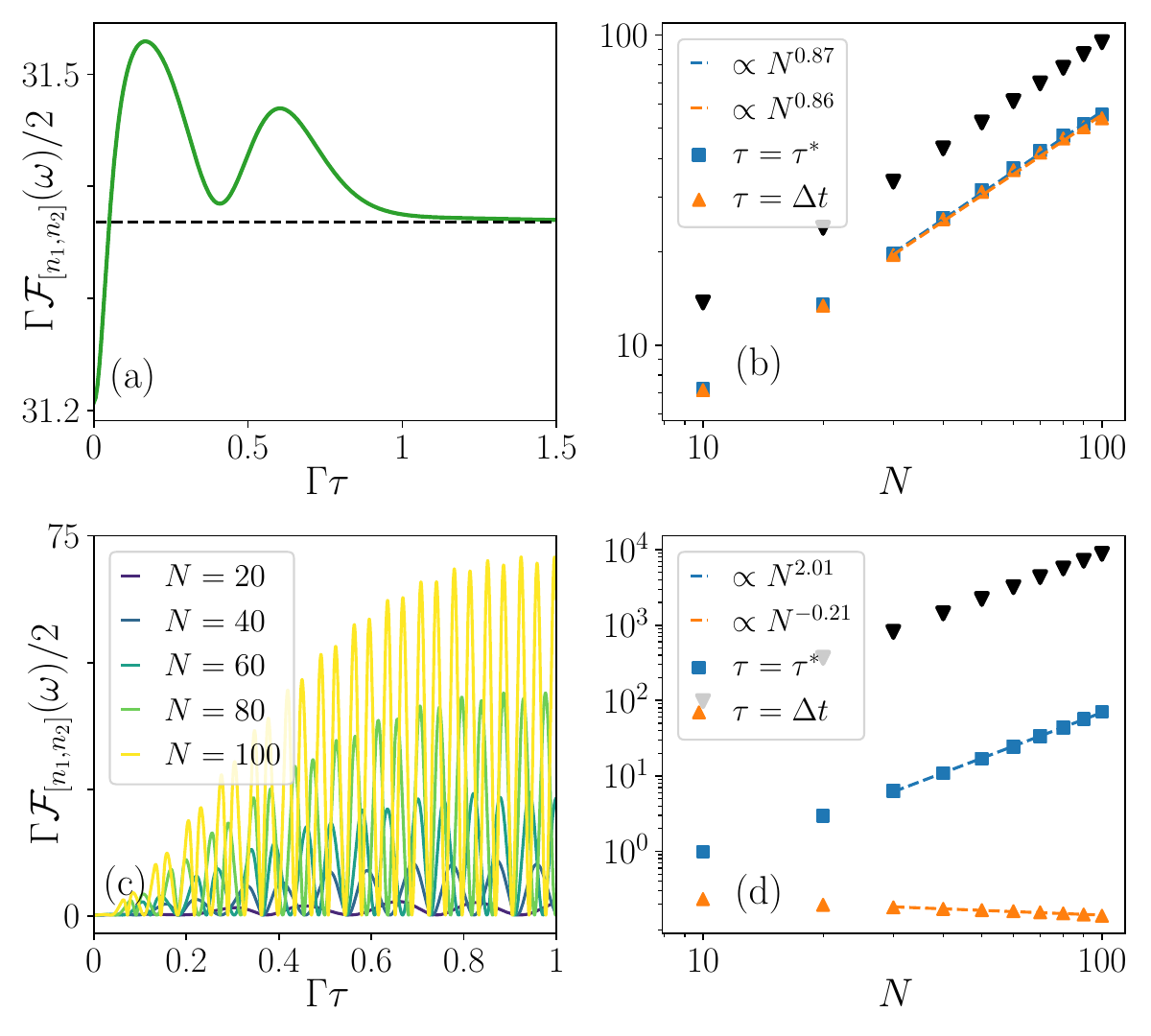}
 \caption{{\bf QFI per unit of time for the two time-bin modes state.} (a) Green solid line:   $\mathcal{F}_{[n_1,n_2]}(\omega)/2$ for $\omega=\omega_\mathrm{c}$, $N=50$, and varying the time between the modes $\tau=(n_2-n_1-1)\Delta t$. Black dashed line:  $\mathcal{F}_{[n_1]}(\omega)$ for the same case. (b) $\mathcal{F}_{[n_1,n_2]}(\omega)/2$ for $\omega=\omega_\mathrm{c}$ and varying $N$ for two different cases: $n_2=n_1+1$ (orange triangles), and for the optimal $\tau=\tau^*$ (blue squares) when the QFI is maximal. Black triangles correspond to $\mathcal{F}_\mathrm{SE}(\omega)$ for the same parameter values \cite{Cabot2024}. Dashed lines correspond to a fit $\propto N^\alpha$ to the largest system sizes. In panel (a) and (b) we have fixed  $\Gamma\Delta t=10^{-5}$. (c) and (d) Same quantities but for the time-crystal phase $\omega=2\omega_\mathrm{c}$. In this case, in panel (c) various $N$ are chosen. In panel (c) and (d) we have fixed  $\Gamma\Delta t=2.5\cdot 10^{-5}$. }
 \label{fig_QFI2_TD}
\end{figure}

We now consider the QFI for the reduced two time-bin state in the small-$\Delta t$ regime and for $n_1\gg 1$. In the two time-bin case, the interesting results are found when studying the QFI as a function of time [see Fig. \ref{fig_QFI2_TD}]. We omit from the analysis the overdamped regime, $\omega/\omega_\mathrm{c}<1$, since the time-bin modes are in a product state.

When considering two consecutive time-bin modes [i.e. $\tau=(n_2-n_1-1)\Delta t=0$] and for $n_1\gg 1$, the QFI displays a similar pattern when varying $\omega/\omega_\mathrm{c}$ as that of the one time-bin mode case (not shown here). A peculiarity of the two consecutive time-bin mode case is that the QFI displays  a slight subadditive behavior, i.e. $\mathcal{F}_{[n_1,n_1+1]}(\omega)<2\mathcal{F}_{[n_1]}(\omega)$ [see Fig. \ref{fig_QFI2_TD} (a)]. Subadditive behavior has been reported for other systems \cite{Radaelli2023}, and it results from correlations acting in a detrimental way for parameter estimation. Nevertheless, the opposite behavior is also observed in our system at the critical point and in the oscillatory regime, where the QFI displays a maximum value for an optimal sensing time $\tau^*$ that depends on the parameter values, but not significantly on system size. This is shown in  Fig. \ref{fig_QFI2_TD} for $\omega/\omega_\mathrm{c}=1$ and $\omega/\omega_\mathrm{c}=2$.

In Fig. \ref{fig_QFI2_TD} (a) we show the QFI per unit of time and varying $\tau$ (green solid line)  for $\omega=\omega_\mathrm{c}$ and $N=50$.  We also show the corresponding $\mathcal{F}_{[n_1]}(\omega)$ with a black dashed line. Subadditivity is observed for small $\tau$, while for large $\tau$ we recover $\mathcal{F}_{[n_1,n_2]}(\omega)\to 2\mathcal{F}_{[n_1]}(\omega)$ when $n_2\gg 1$, since the two time-bin modes become completely uncorrelated. A maximum for the QFI is found at time $\tau^*$.  In Fig. \ref{fig_QFI2_TD} (b) we compare this maximum QFI (blue squares) with the one for two consecutive time bins (orange triangles) varying $N$. We observe that, for $\omega=\omega_\mathrm{c}$ they display almost the same scaling, and thus sensing at the optimal time only offers a small constant gain. 

In   Fig. \ref{fig_QFI2_TD} (c) and (d) we consider the same quantities in the time-crystal phase $\omega/\omega_\mathrm{c}=2$. In panel (c) we plot $\mathcal{F}_{[n_1,n_2]}(\omega)$ varying $\tau$ and $N$. Oscillations occur varying $\tau$ and displaying a frequency that is twice that of the magnetization dynamics, which is $\Omega=\sqrt{\omega^2-\omega_\mathrm{c}^2}$ \cite{Carmichael1980}. While not visible in the plot,  subadditive behavior is also observed for small $\tau$, while for large $\tau$ the two time-bin mode reduced state factorizes. The QFI develops a maximum around $\Gamma \tau^*\sim 1$, the specific time depending on $N$ due to the dependence of the oscillation frequency on system size. The most interesting result regards the behavior of the QFI varying $N$ in the small-$\Delta t$ regime, see Fig. \ref{fig_QFI2_TD} (d). Here we find that at the optimal sensing time, the QFI displays the scaling $\mathcal{F}_{[n_1,n_2]}(\omega)\propto N^2$, recovering the same dependence with $N$ as for the full  system-emission joint state, $\mathcal{F}_\mathrm{SE}(\omega)$ \cite{Cabot2024},  and shown here in black triangles. This suggests that that the collective effects enhancing the QFI of the full system-emission joint state  begin to be present when probing the field at just two times and in the small-$\Delta t$ regime.

\subsection{Sensitivity of the full protocol and asymptotic limit}\label{sec_ultimate_limit}

\begin{figure}[t!]
 \centering
 \includegraphics[width=1\columnwidth]{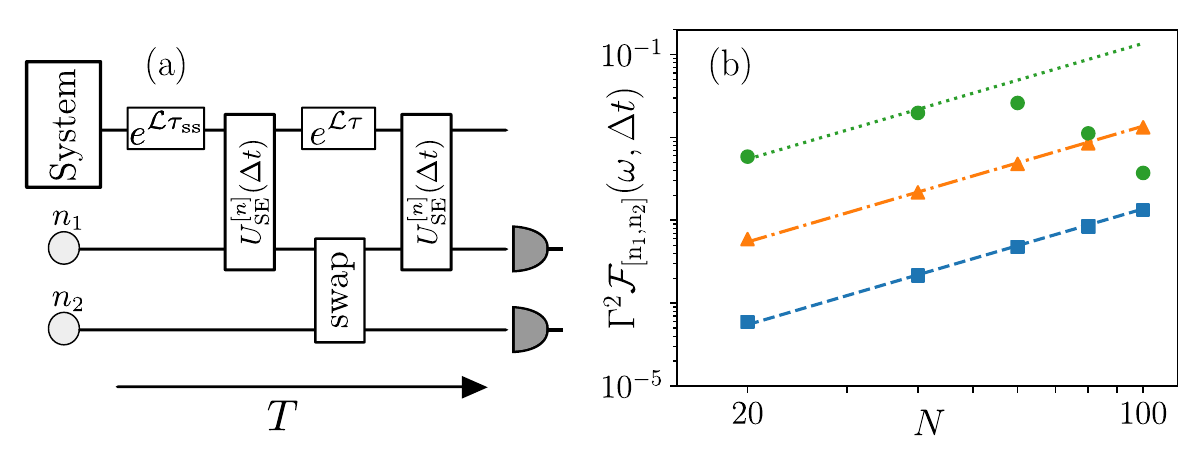}
 \caption{{\bf Sensing protocol beyond the small-$\Delta t$ regime.} (a) Schematic illustration of the full sensing protocol taking a total time duration $T$. (b) QFI of the sensing protocol of panel (a) optimized with respect to $\tau$ for two time-bin modes that have interacted with the system for a time $\Delta t$,  with $\omega/\omega_\mathrm{c}=2$ and varying system size $N$. Blue squares correspond to $\Gamma\Delta t=10^{-5}$, orange triangles to $\Gamma\Delta t=10^{-4}$, and green circles $\Gamma\Delta t=10^{-3}$. Broken lines correspond to reference lines displaying a $N^2$ scaling. For the largest $\Delta t$ value, the large $N$ points deviate from the $N^2$ displayed in the small-$\Delta t$ regime.}
 \label{fig_scaling_deviation}
\end{figure}

We now compare the sensitivity of the one and two time-bin modes protocols with the bounds obtained in Refs. \cite{Montenegro2023,o2025quantum} applicable to our system when the output is not monitored with unit efficiency. Schematically, our sensing protocols can be visualised as in Fig. \ref{fig_scaling_deviation} (a). The protocol involving only one measurement ends after the first unitary gate depicted in Fig. \ref{fig_scaling_deviation} (a), while the protocol involving two time-bin modes ends after the second unitary gate. Essentially, the total time to perform the protocol is given by 
\begin{equation}\label{time_protocol}
T\approx\tau_\mathrm{ss}+\tau,
\end{equation}
where $\tau_\mathrm{ss}$ is the time to reach the stationary state, which for our system (besides the stationary phase) is roughly independent of $N$ and of the same order of magnitude as $1/\Gamma$. Notice that  in Eq. (\ref{time_protocol}) we have neglected the time duration of the unitary gates since $\Delta t\ll \tau_\mathrm{ss}$. The sensitivity of our protocols is ultimately upper bounded by \cite{Montenegro2023,o2025quantum}
\begin{equation}\label{ultimate_bound}
\mathcal{F}_\mathrm{[n_1]}(\omega,\Delta t), \,\,\mathcal{F}_\mathrm{[n_1,n_2]}(\omega,\Delta t)\le \frac{T}{\Gamma}.
\end{equation}
This general bound has been derived in Refs. \cite{Montenegro2023,o2025quantum} using the techniques of noisy quantum metrology \cite{Sekatski2017,demkowicz2017adaptative,zhou2018achieving,wan2022bounds}. It arises because the Hamiltonian term encoding the parameter of interest, $\omega S_\mathrm{x}$, lies in the same  span as the unmonitored decay channel responsible for the noise, $\sqrt{\Gamma} S_-$. Under these conditions, the QFI cannot grow faster than linearly with $T$ \cite{demkowicz2017adaptative}, and it may also restrict the scaling with $N$, as happens in the present scenario \cite{Montenegro2023,o2025quantum}.
This bound applies to protocols exploiting the long-time dynamics and monitoring efficiency tending to zero. These are precisely the conditions under which our protocols operate, since on a time of the order of $\tau_\mathrm{ss}$ the emission field is only probed once or twice.

When recovering the factor $\Gamma\Delta t$ in the results presented in Figs. \ref{fig_QFI_one_ancilla} and \ref{fig_QFI2_TD}, we observe that this bound is respected in all cases, since both  $\mathcal{F}_\mathrm{[n_1]}(\omega,\Delta t),\mathcal{F}_\mathrm{[n_1,n_2]}(\omega,\Delta t)\ll 1/ \Gamma^2$, while the bound, given that $T\sim 1/\Gamma$, is of order $1/\Gamma^2$. An interesting question is whether by increasing sufficiently $N$, the smallness of the factor $\Gamma \Delta t$ can be compensated. The answer is negative, as the results presented in the previous subsections attain the small-$\Delta t$ regime in which Eqs. (\ref{linear_state}) and (\ref{analytical_two_ancilla}) are valid. In fact, when keeping a small but fixed $\Delta t$, and increasing $N$, the observed scaling behaviors are eventually lost due to the small-$\Delta t$ physics of Eqs. (\ref{linear_state}) and (\ref{analytical_two_ancilla}) ceasing to be valid. This is clearly illustrated in Fig. \ref{fig_scaling_deviation} (b), in which we plot the maximal $\mathcal{F}_\mathrm{[n_1,n_2]}(\omega,\Delta t)$ in the time-crystal regime for different fixed values of $\Delta t$ and increasing $N$. For the smallest values of $\Delta t$, the results still follow the expected $N^2$ scaling. However, for the largest value considered, $\Gamma\Delta t=10^{-3}$, clear deviations from this behavior appear because the small-$\Delta t$ condition $N^2\Gamma\Delta t\ll 1$ is no longer satisfied for the system sizes under study.

\section{Sensitivity of two-time measurements}\label{Sec_measurements}

\subsection{Measurements on the time-bin modes reduced state}

In this section, we show that the QFI of two-time bin modes can be effectively exploited by correlated counting measurements at two different times. First we recall that the quantum Cramér-Rao bound provides a link between the Fisher information and the lowest variance that can be achieved by performing measurements on these reduced states \cite{Braun2018}. In particular, when measuring $K$ times an observable $\hat{A}$ on these reduced states, with $K\gg1$, we obtain an estimation  of a parameter $g$ with the following variance:
\begin{equation}
\Delta g(K)|_{\hat{A}} = \frac{\Delta g|_{\hat{A}}}{K},
\end{equation}
where 
\begin{equation}
\Delta g|_{\hat{A}}=(\langle\hat{A}^2\rangle-\langle\hat{A}\rangle^2)\,\,\bigg|\frac{\partial\langle \hat{A}\rangle}{\partial g}\bigg|^{-2},
\end{equation}
and expected values $\langle \dots \rangle$ are taken with respect to $\hat{\mu}_{[n_1]}$ or $\hat{\mu}_{[n_1,n_2]}$. We refer to $\Delta g|_{\hat{A}}$ as the {\it estimation error}. The quantum Cramér-Rao bound then reads as:
\begin{equation}\label{QRB_timebins}
\Delta g(K)|_{\hat{A}}\geq \frac{1}{K \Delta t \mathcal{F}_{q}(g)},    
\end{equation}
where $q=[n_1]$ or $q=[n_1,n_2]$, depending on whether the observable $\hat{A}$ refers to one or two time-bin modes, respectively. Here, we have also assumed the small-$\Delta t$ regime in which a linear dependence of the QFI with $\Delta t$ is found.

\subsection{Measurement at the output of an interferometer}

We now proceed to analyze measurement schemes that can be implemented with the Mach-Zehnder interferometer depicted in Fig. \ref{fig_cartoon}. We assume 50:50 beam splitters with a $\pi/2$ phase between reflection an transmission. The input of arm '0' is the light emitted  by the system through the collective channel, while the one of arm '1' is the vacuum. We then place one photon counter at each of the output arms '4' and '5'. The optical path difference is selected in order to match the time difference $\tau=(n_2-n_1-1)\Delta t$ that we are interested in probing. We analyze the performance of photon counting measurements at each of the output arms, and also of the subtraction of both counting signals. This gives us access to the observables:
\begin{equation}
\begin{split}
\hat{N}_\mathrm{4(5)}=\hat{a}_{4(5)}^\dagger \hat{a}_{4(5)},\quad \hat{N}_\mathrm{d}&=\hat{N}_5-  \hat{N}_4,
\end{split}
\end{equation}
where $\hat{a}_{4,5}$ are the annihilation operators for the light field at the output arms. A detailed description of the interferometer in terms of time-bin modes is given in Appendix \ref{appendix_interferometer}, including the expression of $\hat{a}_{4,5}$ in terms of the input modes. The statistics of $\hat{N}_\mathrm{4,5,d}$ can be fully characterized with the two time-bin modes reduced state $\hat{\mu}_{[n_1,n_2]}$. Their expected values with respect to this state read:
\begin{equation}
\begin{split}
\langle\hat{N}_\mathrm{d}\rangle&=\frac{1}{2}\big[\langle b^\dagger_{[n_1]}b_{[n_2]}\rangle e^{-i\Delta\phi}+\langle b^\dagger_{[n_2]}b_{[n_1]}\rangle e^{i\Delta\phi} \big],\\
\langle\hat{N}_{4,5}\rangle&=\frac{1}{4}\big[\langle b^\dagger_{[n_1]}b_{[n_1]}\rangle+\langle b^\dagger_{[n_2]}b_{[n_2]}\rangle\mp 2\langle\hat{N}_\mathrm{d}\rangle\big],
\end{split}
\end{equation}
where the minus sign in the second equation corresponds to $\hat{N}_4$, while the plus to $\hat{N}_5$. Moreover, we have defined the phase difference $\Delta \phi=\omega_0\tau$. Varying the optical length of the two paths in Fig. \ref{fig_cartoon} (b) results in an interference pattern for the number of counts. This pattern has two very distinct contributions: a very fast one due to $\Delta\phi$ with characteristic scale $\omega_0^{-1}$, and a slow envelope of dynamical origin and with characteristic scale   $\omega^{-1}$. Because of our coarse-grained description, i.e. $\omega_0\gg \omega$, we can decouple these two contributions and fix one value of $\Delta\phi$ to a good approximation, which we take $\Delta\phi=0$ for convenience.

\begin{figure}[t!]
 \centering
 \includegraphics[width=1\columnwidth]{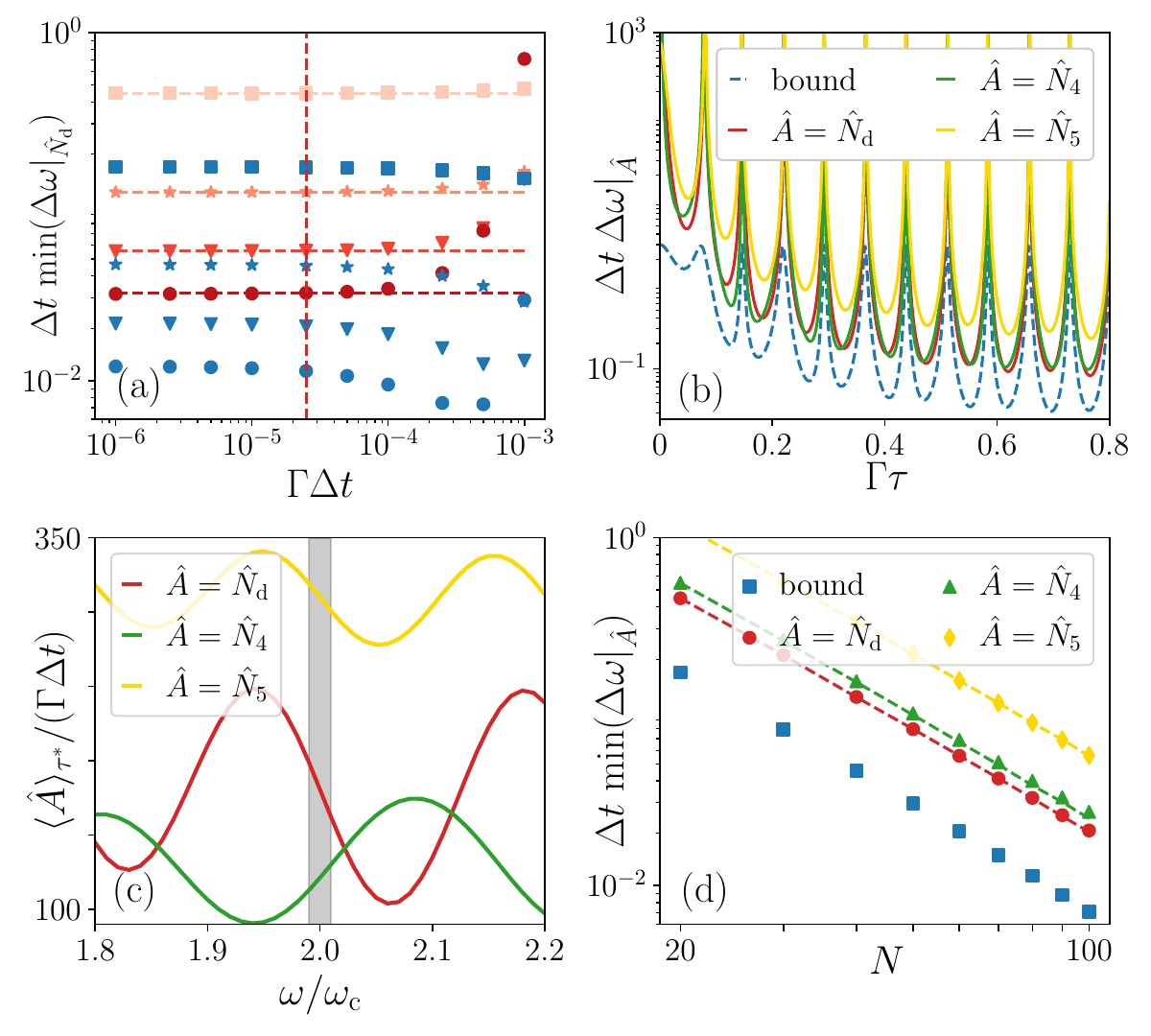}
 \caption{{\bf Characterization of photon counting measurements after the interferometer.}   (a) Red symbols: estimation error $\Delta\omega|_{\hat{N}_\mathrm{d}}$ at the optimal time $\tau^*$ varying $\Gamma \Delta t $ and $N$ ($N=20$ squares, $N=40$ stars, $N=60$ triangles, $N=80$ circles). Dashed red lines correspond to the approximate results of Eq. (\ref{equation_error_analytic}), obtained calculating the two-time correlation functions with Eq. (\ref{ME}). Blue symbols correspond to the fundamental bound for the corresponding parameters. (b) Solid lines: $\Delta\omega|_{\hat{A}}$ as a function of the time between the time-bin modes $\tau$ and $N=50$, for the three considered measurements, $\hat{A}$. Blue dashed line: fundamental bound on sensitivity given by the inverse of the two time-bin QFI per unit of time. (c) Expected values of the different counting measurements at their corresponding optimal time $\tau^*$ for sensing around $\omega/\omega_\mathrm{c}=2$ (shaded region). (d) Estimation error at the optimal sensing time $\tau^*$ for each of the three counting measurements and varying system size. The dashed lines correspond to the approximate results of Eq. (\ref{equation_error_analytic}). The chosen $\Gamma \Delta t=2.5\cdot 10^{-5}$ is indicated as a red-dashed vertical line in panels (a). The results of panels (a), (b) and (d) correspond to $\omega=2\omega_\mathrm{c}$.}
 \label{fig_measurements}
\end{figure}

In the small-$\Delta t$ regime, we can use the approximate expression for the  two time-bin modes reduced state  [Eq. (\ref{analytical_two_ancilla})] to calculate these expected values and the corresponding estimation errors. Probing the long-time dynamics ($n_1\gg1$), their leading contributions read:
\begin{equation}\label{equation_error_analytic}
\begin{split}
\Delta \omega|_{\hat{N}_\mathrm{d}}&\approx
\frac{1}{\Gamma\Delta t}\frac{\langle S_+ S_-\rangle_\mathrm{ss}}{|\partial_\omega\text{Re}[\langle S_+(\tau)S_-\rangle_\mathrm{ss}]|^2},\\
\Delta \omega|_{\hat{N}_\mathrm{4,5}}&\approx
\frac{2}{\Gamma\Delta t}\frac{\langle S_+ S_-\rangle_\mathrm{ss}\mp\text{Re}[\langle S_+(\tau)S_-\rangle_\mathrm{ss}]}{|\partial_\omega(\langle S_+ S_-\rangle_\mathrm{ss}\mp\text{Re}[\langle S_+(\tau)S_-\rangle_\mathrm{ss}])|^2},
\end{split}
\end{equation}
We benchmark this expression with the exact dynamics in Fig. \ref{fig_measurements} (a), for the case of $\hat{N}_\mathrm{d}$, while the cases of $\hat{N}_{4,5}$ are presented in Appendix \ref{appendix_interferometer}. As we can see, when $N\Gamma\Delta t$ becomes small enough, the color points (exact dynamics) converge to the dashed lines (approximate solution). This convergence indicates that for the considered system sizes $N$, $\Delta t$ is small enough such that the small-$\Delta t$ expansion is valid. In this case, we choose the value of the estimation error at the optimal sensing time $\tau^*$ at which the estimation error is minimized. We also compare it to the fundamental bound given by the QFI at the optimal sensing time (blue points), finding that this kind of measurement is close to optimal (roughly a factor 3 times the bound).

In Fig. \ref{fig_measurements} (b), we show that the estimation errors (solid lines)  display a similar temporal pattern in $\tau$ as the inverse of the QFI (blue dashed line). Local minima are displayed separated by approximately half the mean-field period of oscillation. Both the bound and the estimation errors display an optimal sensing time $\tau^*$ at which they are minimized. Nevertheless, it is not crucial for the protocol to be tuned to this optimal time, but rather to select a $\tau$ close to one of the many minima, as they display similar estimation errors. While near the local minima the bound and estimation errors display similar values, in between the minima the latter display values orders of magnitude larger. This variability results from a different susceptibility of $\langle S_+(\tau)S_-\rangle_\mathrm{ss}$  to changes in $\omega$ within each period of the sinusoidal oscillations. 

In Fig. \ref{fig_measurements} (c), we fix $\tau$ to the optimal time for $\omega/\omega_\mathrm{c}=2$ and each of the measurements, and we study how their expected values change with the Rabi frequency. This allows us to understand how the sensing protocols work around a particular value of $\omega/\omega_\mathrm{c}$ (shadowed region). We observe that the optimal time  corresponds (approximately) to the point of maximum derivative of the sinusoidal pattern. Notice that a value of $\langle \hat{N}_\mathrm{d,4,5}\rangle$  is not generally associated to a unique Rabi frequency. As a consequence, these protocols can be used to sense small perturbations around a previously calibrated value of $\omega/\omega_\mathrm{c}$.

In Fig. \ref{fig_measurements} (d), we  analyze in more detail the behavior with $N$ of the estimation error for the different measurements at their optimal sensing times. The color points correspond to the exact dynamics while the dashed lines to the results obtained from Eq. (\ref{equation_error_analytic}). They display similar scaling laws, in between $N^{-1.89}$ and $N^{-1.95}$, which are close to the one of the bound. While the protocol based on photon subtraction is the most sensitive, all of them display similar values. Therefore, we conclude that measurement protocols based on photon counting of the output of the interferometer provide a way to efficiently exploit the sensitivity of two-time measurements in the time-crystal phase.

From the results presented in Fig. \ref{fig_measurements} it becomes clear that the collective oscillations play a fundamental role in the sensitivity of the protocol. The presence of an optimal sensing time and the enhanced scaling with system size can be understood from the properties of $\langle S_+(\tau)S_-\rangle_\mathrm{ss}$ and its derivative with $\omega$. In fact, we can gain understanding  from the following approximate expression for the derivative of two-time correlations with respect to $\omega$ (see Appendix \ref{appendix_ansatz} for more details):
\begin{equation}\label{ansatz_derivative}
\partial_\omega\langle S_+(\tau)S_-\rangle_\mathrm{ss}\approx  -\frac{I_\mathrm{inc} \tau \partial_\omega \Omega}{2\Gamma}\sin \Omega\tau e^{-\Gamma_1 \tau} 
\end{equation}
where $\Gamma_1\approx \Gamma$ for $\omega/\omega_\mathrm{c}=2$, and the incoherent stationary intensity is defined as $I_\mathrm{inc}=\Gamma(\langle S_+S_-\rangle_\mathrm{ss}-\langle S_+\rangle_\mathrm{ss}\langle S_-\rangle_\mathrm{ss})$. This formula has a maximum close to $\tau\sim\Gamma$, in accordance to our observations. Importantly, the only term proportional to $N^2$ is the incoherent intensity. Thus, the observed $N^2$ enhancement results from atom-atom correlations that build up in the oscillatory phase and which lead to $I_\mathrm{inc}\propto N^2$ (see Ref. \cite{Carmichael1980} for a characterization of these atom-atom correlations).

\section{Local decay as additional decay channel}\label{Sec_localdecay}

Until now, we have considered the system to display just one collective decay channel. However, in practice, one might find more decay channels that remain typically unmonitored. Here we address this general problem by considering the effects of local spontaneous emissions  on the sensitivity of the time-crystal phase, and assuming that these channels are not monitored. Local decay is described by the following terms:
\begin{equation}\label{ME_loc_dec}
\mathcal{L}_\mathrm{loc}{\rho}=\gamma\sum_{j=1}^N\big( \sigma_-^{(j)}\rho\sigma_+^{(j)}-\frac{1}{2}\{\sigma_+^{(j)}\sigma_-^{(j)},\rho\}\big),    
\end{equation}
such that the master equation for the system reduced density matrix becomes $\partial_t \rho=(\mathcal{L}+\mathcal{L}_\mathrm{loc}){\rho}$. The time-bin modes implementing the local decay channels give rise to a Kraus map that in the small-$\Delta t$ regime corresponds to $e^{\mathcal{L}_\mathrm{loc}\Delta t}$ (as they remain unmonitored),  which can be efficiently implemented exploiting the permutation symmetry of the system \cite{Chase2008,Baragiola2010}.

\begin{figure}[t!]
 \centering
 \includegraphics[width=1\columnwidth]{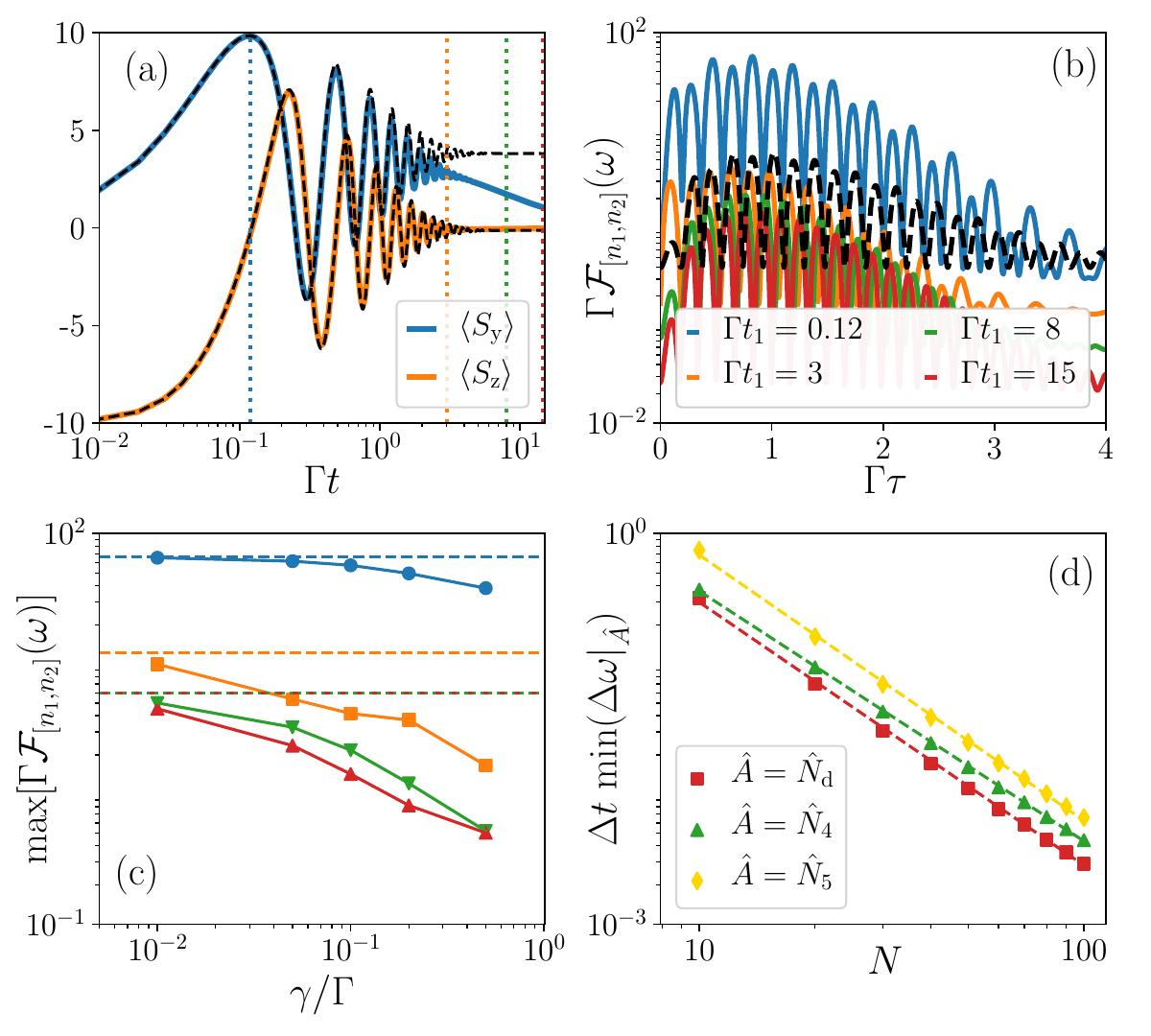}
 \caption{{\bf Impact of local decay.}   (a) Dynamics of the collective magnetizations for $\omega/\omega_\mathrm{c}=2$, $N=20$ $\gamma/\Gamma=0.1$ and all atoms initially in  the ground state. Black dashed lines correspond to the same case without local decay,  $\gamma=0$. (b) QFI per unit of time for the two time-bin mode reduced state varying $\tau$ and for the same parameters as in (a). Different lines correspond to different values of $\Gamma t_1$ at which the first time-bin mode interacts with the system. Black dashed line corresponds to the two time-bin QFI in the stationary state without local decay. (c) Maximum QFI of the two time-bin reduced state varying $\gamma/\Gamma$ for $\omega/\omega_\mathrm{c}=2$, $N=20$ and the same $\Gamma t_1$'s as in panel (b) (indicated by the same color code). Dashed lines display the QFI for $\gamma=0$ and the corresponding $\Gamma t_1$. In panels (b) and (c) we have fixed $\Gamma\Delta t=2.5\cdot10^{-5}$.} (d) Minimum estimation errors with respect to $\tau$, and taking $t_1$ such that it coincides with the first maximum of $\langle S_\mathrm{y}\rangle$ (blue lines and points in the other figures) varying $N$ and for $\gamma/\Gamma=0.1$. The results of this panel are obtained from Eq. (\ref{equation_error_analytic_local_decay}), derived in Appendix \ref{Appendix_local_decay}. The dashed lines correspond to fits of the type $\propto N^{-\alpha}$ with $\alpha=(2.04,1.92,2.04)$ for $\hat{N}_\mathrm{d}$, $\hat{N}_4$ and $\hat{N}_5$, respectively. In all cases we take as  initial condition all atoms in the ground state.
 \label{fig_local_decay}
\end{figure}

In Fig. \ref{fig_local_decay} (b), we show the two time-bin mode QFI per unit of time varying $\tau$ and for different values of $t_1=n_1\Delta t$ with $N=20$ and $\gamma/\Gamma=0.1$. The considered values of $t_1$ are displayed as vertical dashed lines in panel (a).  We observe that the QFI attains its largest values for the smallest $\Gamma t_1$. The dynamical behavior of the QFI for small $\Gamma\tau$ (until $\Gamma \tau \sim 3$) is similar to that in the absence of local losses (black dashed line). For larger $\Gamma \tau$ the oscillations and magnitude displayed by the QFI are attenuated due to the effects of local decay. In all cases the QFI displays an optimal sensing time $\tau^*$ at which it assumes the maximum value. In Fig \ref{fig_local_decay} (c), we analyze how the maximum of the QFI varies with the local decay strength $\gamma/\Gamma$, and for different values of $\Gamma t_1$. For comparison we plot the respective maximum of the QFI for $\gamma=0$ and at the corresponding $\Gamma t_1$ in color dashed lines. As we increase $\gamma/\Gamma$ the maximum QFI diminishes. The effects of local decay are smaller for the smallest values of $\Gamma t_1$. This follows from the metastable character of the collective oscillations. Instead,  for the largest $\Gamma t_1$ and $\gamma/\Gamma$, the time-bin modes are essentially resolving the stationary state in the presence of local losses, which displays a smaller value of the  QFI.

We now analyze the effects of local decay on the specific sensing protocols based on photon counting at the output of the interferometer (see Fig. \ref{fig_cartoon}). Within the metastable transient, the dynamics in $\tau$  of the estimation error are qualitatively similar to the case without local losses, also closely following the fundamental bound (not shown here). The approximate formulas of Eq. (\ref{equation_error_analytic}) work well also in the presence of local decay and in the small-$\Delta t$ regime (see Appendix \ref{Appendix_local_decay} for more details). Notice that one has to evaluate the expectation values and two-time correlations in Eq. (\ref{equation_error_analytic}) using the total Liouvillian $\mathcal{L}+\mathcal{L}_\mathrm{loc}$, and replacing the stationary time by the corresponding $\Gamma t_1$ [see Eq. (\ref{equation_error_analytic_local_decay})]. In Fig. \ref{fig_local_decay} (d) we analyze the effects of increasing $N$ on the estimation error at the optimal sensing time, $\Gamma\tau^*\sim 1$, and using Eq. (\ref{equation_error_analytic_local_decay}) that includes local decay and it is valid in the small-$\Delta t$ regime. We focus on the results obtained fixing $\Gamma t_1$ to the first oscillation maximum of $\langle S_\mathrm{y}\rangle$ ($\Gamma t_1=0.12$ in the case of $N=20$). Similarly to the case without local decay, the measurement of $\hat{N}_\mathrm{d}$ is the one providing more sensitivity, although $\hat{N}_{4,5}$ display similar values. Remarkably, we observe an  approximate quadratic scaling with particle number, $N^{-2}$, in all cases recovering  the one observed in the absence of local decay. Notice that the results of Fig. \ref{fig_local_decay} (d) have been obtained using the small-$\Delta t$ expansion for the two time-bins reduced state, and therefore the observed quadratic scaling is only maintained up to a certain size that depends on the value of the fixed $\Delta t$. This scaling is also sensitive to the time $\Gamma t_1$, finding that at later times the scaling is no longer quadratic (see Appendix \ref{Appendix_local_decay}). In this sense, in the presence of local losses, one could exploit the quadratic scaling in $N$ using a pulsed scheme in which the system is periodically reset to the ground state in periods of the order of the optimal sensing time.

\section{Discussion and conclusions}\label{Sec_conclusions}

We have analyzed the problem of parameter estimation using the emitted light of a system displaying a nonequilibrium phase transition between a stationary phase and a time-crystal phase (the BTC). In particular, we have addressed the question of whether these collective phenomena can be observed and harnessed for parameter estimation when measuring the emission field solely at one time or at two different times. Interestingly, we have found that at the nonequilibrium phase transition and in the short (probing) time limit, measuring the emission field at one time is already enough to resolve collective effects, which manifest as a transient enhancement of the QFI with system size.  In contrast, within the time‑crystal phase, sensitivity enhanced by collective phenomena arises only when measurements can resolve at least two‑time correlations of the emission field.  In this regime, we observe a transient $N^2$ scaling of the QFI, and we have shown that it can be accessed through intensity measurements at the output of a Mach–Zehnder interferometer. Our analysis also provides a clear picture of the resource that is enhancing the sensitivity in the time-crystal phase. In particular, we have shown that efficient measurements probe the emission field around the maximum gradient point of the collective oscillations with respect to changes in the parameter $\omega$ [see Fig. \ref{fig_measurements}], and that the transient $N^2$ sensitivity is rooted in the build up of atom-atom correlations in the oscillatory phase [see Eq. (\ref{ansatz_derivative})]. 

Moreover, we have compared the sensitivity of these protocols with the ultimate size independent bound obtained using tools from noisy quantum metrology \cite{Montenegro2023,o2025quantum}. We have found that the transient system size scalings observed at the nonequilibrium phase transition and within the time‑crystal phase eventually disappear once $N$ becomes large enough that deviations from the small-$\Delta t$ probing regime become significant for a fixed probing time $\Delta t$. While our discussion has focused on measurements of the emitted field, where $\Delta t$ may not be directly controllable, our results also apply to scenarios in which a subset of two‑level systems within the ensemble is used to probe the rest, as it can be naturally implemented in trapped‑ion platforms \cite{Shankar2017}. In such cases, $\Delta t$ is tunable, making the distinction between the small-$\Delta t$ regime and its breakdown particularly relevant.

Finally, we have addressed the effects of local losses on our sensing protocols. We have found that for losses up to the order of $\gamma/\Gamma\sim 0.1$ a metastable oscillatory transient persists, in which the estimation errors for counting measurements still display the (transient) $N^{-2}$ enhancement in the small-$\Delta t$ regime. In this case, we propose the use of pulsed schemes in order to gather statistics about this interesting transient. We remark that the values of $\gamma/\Gamma$ considered here are still comparatively small to those found in cavity QED setups in which collective atomic physics can be observed (see e.g. discussion in Ref. \cite{Xu2015} or atom-cavity cooperativity values reported in, e.g., Refs. \cite{Norcia2016b,Norcia2016c}). However, in these scenarios, collective effects are enhanced by  resorting on much larger atomic numbers such that the so-called strong collective coupling regime is reached. In this sense, our analysis highlights that the collective effects still enhance the sensitivity when the collective decay channel is the dominant one, and that we can overcome the presence of undesired decay channels by focusing on the collective effects present in the transient dynamics.

\section{Acknowledgements} 

We thank F. Albarelli for interesting discussions. AC acknowledges support from the Deutsche Forschungsgemeinschaft (DFG, German Research Foundation) through the Walter Benjamin programme, Grant No. 519847240 and from both the Spanish Ministerio de Ciencia, Innovación y Universidades and  Universitat de les Illes Balears through the Beatriz
Galindo programme (BG24/00134). FC~is indebted to the Baden-W\"urttemberg Stiftung for the financial support of this research project by the Eliteprogramme for Postdocs. We acknowledge the use of Qutip python library \cite{Qutip1,Qutip2}. We acknowledge funding from the Deutsche Forschungsgemeinschaft (DFG, German Research Foundation) through the Research Unit FOR 5413/1, Grant No.~465199066. We acknowledge support by the state of Baden-Württemberg through bwHPC and the German Research Foundation (DFG) through grant no INST 40/575-1 FUGG (JUSTUS 2 cluster). This work was supported by the QuantERA II programme (project CoQuaDis, DFG Grant No. 532763411) that has received funding from the EU H2020 research and innovation programme under GA No. 101017733. 
\appendix

\section{Implementation of the discrete dynamics of the system plus one or two time-bin modes}\label{Appendix_collision_model}

{\it System plus one time-bin mode. --} We first assume that we are interested in keeping the state of the $n_1$-th time-bin mode.  After the interaction of the system with this time-bin mode, their joint reduced state is:
\begin{equation}
\hat{\varrho}_{[n_1]}(n_1\Delta t)=e^{-iH_{[n_1]}\Delta t}\big(\mathcal{E}^{n_1-1}\rho(0)\big)\otimes |0_{n_1}\rangle \langle 0_{n_1}|  e^{iH_{[n_1]}\Delta t}.   
\end{equation}
Here we have defined the Hamiltonian between the system and the $n_1$-th time-bin:
\begin{equation}
H_{[n_1]}=\omega S_\mathrm{x}+i\sqrt{\frac{\Gamma}{\Delta t}}[S_-{b}_{[n_1]}^\dagger-S_+{b}_{[n_1]}],    
\end{equation}
whose action can be used to implement the input-output dynamics presented in Sec. \ref{Sec_discrete} \cite{Ciccarello2022}. In order to further advance the dynamics  keeping track of time-bin $n_1$, we have to make use of the extended Hamiltonian: 
\begin{equation}\label{Ham_CM_1}
H_{[n_1,n]}=\omega S_\mathrm{x}+i\sqrt{\frac{\Gamma}{\Delta t}}[S_-\otimes \mathbb{I}_2\otimes b_{[n]}^\dagger-S_+\otimes \mathbb{I}_2\otimes b_{[n]}].   
\end{equation}
where $\mathbb{I}_\mathrm{2}$ is the identity acting over the time-bin $n_1$, while the label $n$ denotes a subsequent time-bin mode. Notice that after Eq. (\ref{unitary_n}) we have neglected two photon transitions, and thus we effectively treat the time-bin degrees of freedom as two-level systems. This amounts to the identification $b_{[n]}\to\hat{\sigma}_-$, $b^\dagger_{[n]}\to\hat{\sigma}_+$, where $\hat{\sigma}_{\alpha}$ ($\alpha=\mathrm{x,y,z},\pm$) are Pauli matrices. If we trace out the subsequent time-bin mode, the reduced state of system and time-bin $n_1$ evolves according to the map:
\begin{equation}\label{CM_1}
\hat{\varrho}_{[n_1]}(n\Delta t)=\mathcal{E}_{[n_1]}\hat{\varrho}_{[n_1]}([n-1]\Delta t)
\end{equation}
defined as:
\begin{equation}
\mathcal{E}_{[n_1]}(\cdot)=K_{0,[n_1]}(\cdot)K_{0,[n_1]}^\dagger+K_{1,[n_1]}(\cdot)K_{1,[n_1]}^\dagger,
\end{equation}
with
\begin{equation}
\begin{split}
K_{0,[n_1]}&=\text{Tr}_{[n]}\{e^{-iH_{[n_1,n]}\Delta t}\big(\mathbb{I}_\mathrm{S}\otimes\mathbb{I}_2\otimes|0_{n}\rangle \langle 0_{n}| \big)\},\\
K_{1,[n_1]}&=\text{Tr}_{[n]}\{e^{-iH_{[n_1,n]}\Delta t}\big(\mathbb{I}_\mathrm{S}\otimes\mathbb{I}_2\otimes|0_{n}\rangle \langle 1_{n}| \big)\}.
\end{split}
\end{equation}
where  $\mathbb{I}_\mathrm{S}$ is the identity for the system. The reduced state for the $n_1$-th time-bin, $\hat{\mu}_{[n_1]}(T)$, is then simply obtained by tracing out the system degrees of freedom.

{\it System plus two time-bin modes. --} We now want to obtain the joint reduced state of the system and two time-bin modes $n_1$ and $n_2$, where $n_2>n_1$. This reduced state just after the interaction with the second mode, i.e. $\hat{\varrho}_{[n_1,n_2]}(n_2\Delta t)$, can be computed using the extended Hamiltonian (\ref{Ham_CM_1}) and Kraus map (\ref{CM_1}), which reads:
\begin{equation}
\begin{split}
\hat{\varrho}_{[n_1,n_2]}(n_2\Delta t)=&    e^{-iH_{[n_2,n_1]}\Delta t}\hat{\varrho}_{[n_1]}\big((n_2-1)\Delta t\big)\\
&\otimes|0_{n_2}\rangle \langle 0_{n_2}|e^{iH_{[n_2,n_1]}\Delta t},
\end{split}
\end{equation}
where
\begin{equation}
\hat{\varrho}_{[n_1]}\big((n_2-1)\Delta t\big)=\mathcal{E}_{[n_1]}^{n_2-n_1-1}  \hat{\varrho}_{[n_1]}(n_1\Delta t). 
\end{equation}
By tracing out the system, we can obtain the state $\hat{\mu}_{[n_1,n_2]}(n_2\Delta t)$. In order to further advance the dynamics of $\hat{\varrho}_{[n_1,n_2]}(n_2\Delta t)$ we  have to extend once more the discrete model to include additional degrees of freedom, similarly to what we have done in Eqs. (\ref{Ham_CM_1}) and (\ref{CM_1}). Nevertheless, this is only necessary if we want to keep more than two time-bin modes, as the information present in the reduced time-bin states, e.g. $\hat{\mu}_{[n_1,n_2]}(T)$, is actually independent of what happens after the last interaction, i.e. it does not change for $T>n_2\Delta t$. In this work, we focus on just one or two time-bin modes.

{\it small-$\Delta t$ regime. --} In the small-$\Delta t$ regime $N^2\Gamma \Delta t \ll 1$, it can be useful to approximate the maps $\mathcal{E}$ and $\mathcal{E}_{[n_1]}$ in terms of the action of master equations. This allows us to obtain the  analytic approximate expressions for the time-bin modes reduced states presented in the main text, i.e. Eqs. (\ref{linear_state}) and (\ref{analytical_two_ancilla}). In particular, the map $\mathcal{E}$ can be well approximated by $e^{\mathcal{L}\Delta t}$, where the Liouvillian is defined in Eq. (\ref{ME}). In turn, the extended map $\mathcal{E}_{[n_1]}$ can be well approximated by $e^{\mathcal{L}_{[n_1]}\Delta t}=e^{\mathcal{L}\Delta t}\otimes\mathbb{I}_2$, i.e.:
\begin{equation}
\begin{split}
\mathcal{L}_{[n_1]}\hat{\varrho}_{[n_1]}=&-i[\omega S_\mathrm{x}\otimes \mathbb{I}_2,\hat{\varrho}_{[n_1]}]+\Gamma \big({S}_-\otimes \mathbb{I}_2\,\hat{\varrho}_{[n_1]}\,{S}_+\otimes \mathbb{I}_2\\
&-\frac{1}{2}\{ {S}_+{S}_-\otimes \mathbb{I}_2,\hat{\varrho}_{[n_1]}\} \big),
\end{split}
\end{equation}
For finite system sizes $N$ we have only one stationary state. The spectrum of $\mathcal{L}$ is given by:
\begin{equation}
\mathcal{L}r_j=\lambda_j r_j,\quad l^\dagger_j \mathcal{L}=\lambda_j l^\dagger_j, \quad \text{Tr}_\mathrm{S}\{l^\dagger_j r_k\}=\delta_{j,k},\quad \rho_\mathrm{ss}=r_0,
\end{equation}
where the eigenvalues are ordered such that $\text{Re}[\lambda_j]\geq \text{Re}[\lambda_k]$ for $j>k$. Then, the spectrum of $\mathcal{L}_{[n_1]}$ is composed of the following tupples for each possible $j$ that combine the eigenmatrices of $\mathcal{L}$ with the basis elements of the time-bin Hilbert space:
\begin{equation}\label{spectrum_L_extended}
\begin{split}
&\{l^\dagger_{j,a}\}_{a=0,1,2,3}=\{l^\dagger_j\otimes \mathbb{I}_2,l^\dagger_j\otimes\hat{\sigma}_\mathrm{x},l^\dagger_j\otimes\hat{\sigma}_\mathrm{y},l^\dagger_j\otimes\hat{\sigma}_\mathrm{z}\},\\
&\{r_{j,a}\}_{a=0,1,2,3}=  \big\{r_j\otimes \frac{\mathbb{I}_2}{2},r_j\otimes\frac{\hat{\sigma}_\mathrm{x}}{2},r_j\otimes\frac{\hat{\sigma}_\mathrm{y}}{2},r_j\otimes\frac{\hat{\sigma}_\mathrm{z}}{2}\big\},\\ &\{\lambda_{j,a}\}_{a=0,1,2,3}=\{\lambda_j,\lambda_j,\lambda_j,\lambda_j\}. 
\end{split}
\end{equation}
Thus the eigenvalues of $\mathcal{L}_{[n_1]}$ are those of $\mathcal{L}$ four-fold degenerate. It follows that in the small-$\Delta t$ regime, the time evolution of the joint system-one-time-bin state can be written as:
\begin{equation}\label{state_L_extended}
\begin{split}
 \hat{\varrho}_{[n_1]}(n\Delta t)&=\sum_{a=0}^3 \text{Tr}_{\mathrm{S},[n_1]}\{l^\dagger_{0,a}\hat{\varrho}_{[n_1]}(n_1\Delta t)\}r_{0,a}\\
 &+  \sum_{j\geq 1} \sum_{a=0}^3 \text{Tr}_{\mathrm{S},[n_1]}\{l^\dagger_{j,a}\hat{\varrho}_{[n_1]}(n_1\Delta t)\}r_{j,a} e^{\lambda_j (n-n_1)\Delta t}.
 \end{split}
\end{equation}
Using that $\text{Tr}_\mathrm{S}\{r_{j\geq1,a}\}=0$, we find that the reduced state of the time-bin only depends on the time $n_1\Delta t$, and it is given by:
\begin{equation}
\hat{\mu}_{[n_1]}=    \sum_{a=0}^3 \text{Tr}_{\mathrm{S},[n_1]}\{l^\dagger_{0,a}\hat{\varrho}_{[n_1]}(n_1\Delta t)\}\text{Tr}_\mathrm{S}\{r_{0,a}\}.
\end{equation}
Similarly, in the stationary state of $\mathcal{L}_{[n_1]}$, correlations between system and time-bin have decayed out resulting in the following (initial condition dependent) stationary state: 
\begin{equation}
\lim_{n\to\infty}    \hat{\varrho}_{[n_1]}(n\Delta t)= \rho_\mathrm{ss}\otimes\hat{\mu}_{[n_1]},
\end{equation}
where $\hat{\mu}_{[n_1]}$ is given by Eq. (\ref{linear_state}). Moreover, from Eqs. (\ref{spectrum_L_extended}) and (\ref{state_L_extended}) it also follows that the dynamics of any observable or multitime correlation that depends only on  system degrees of freedom can be computed just using the system Liouvillian $\mathcal{L}$. This result is used in the derivation of Eq. (\ref{analytical_two_ancilla}), which contain system two-time correlations.

\begin{figure}[t!]
 \centering
 \includegraphics[width=1\columnwidth]{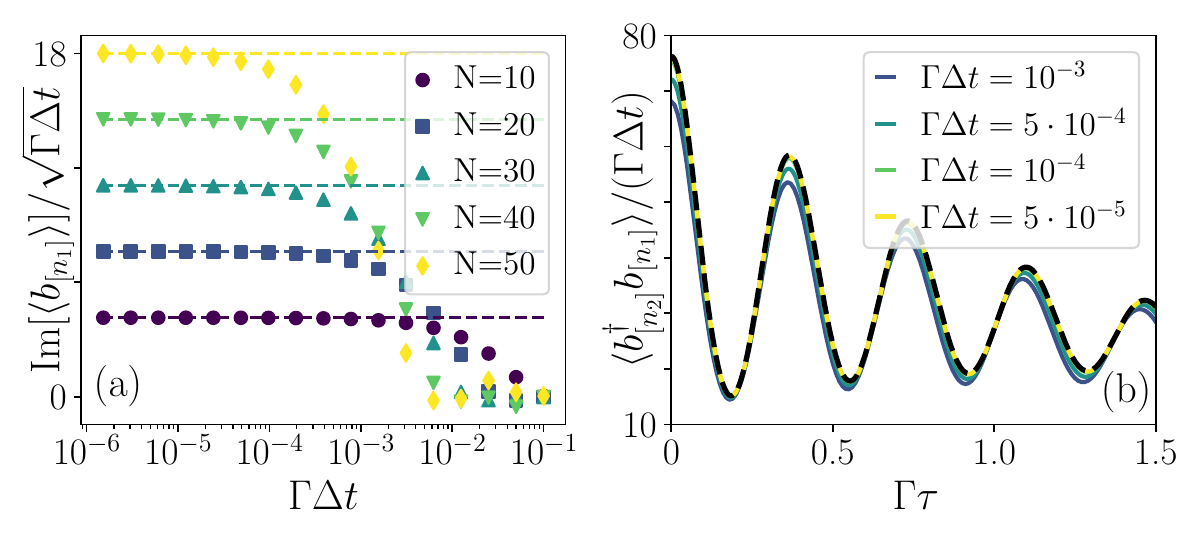}
 \caption{{\bf Time-bin mode observables for small $\Delta t$.} (a) Color points indicate the time-bin expectation value computed using the discrete time evolution varying $\Gamma\Delta t$ and for various $N$. Dashed lines correspond to the approximate result computed from Eqs. (\ref{linear_state}) with  Eq. (\ref{ME}). The results are rescaled by $\sqrt{\Gamma \Delta t}$ and are computed for $n_1$ such that the system is in the stationary state. (b) Color points: two-time bin mode observables for $N=20$, varying the time between time-bins  $\Gamma\tau$ and for different values of  $\Gamma\Delta t$. Dashed black line: result according to Eq.  (\ref{analytical_two_ancilla}) computed with Eq. (\ref{ME}). The results have been scaled by $\Gamma\Delta t$. In both cases $\omega/\omega_\mathrm{c}=2$. }
 \label{fig_convergence_ancillas}
\end{figure}

\textit{Numerical results in the  small-$\Delta t$ regime. --} In Fig. \ref{fig_convergence_ancillas} we show that, as the small-$\Delta t$ regime is approached, observables depending on one-time bin and two-time bins approach the values corresponding to Eqs. (\ref{linear_state}) and (\ref{analytical_two_ancilla}). In particular, in Fig. \ref{fig_convergence_ancillas} (a) we consider one time-bin mode observable for different system sizes $N$ and varying the time $\Delta t$. The results obtained integrating the Kraus map (color points) approach those computed with Eq. (\ref{linear_state}) and the master equation as $\Delta t$ diminishes. We notice that the larger the system size, the smaller needs to be the interaction time for both results to converge. This reflects the fact that the linear expansion in Eq. (\ref{unitary_n}) is valid for $N^2\Gamma\Delta t\ll 1$. In Fig. \ref{fig_convergence_ancillas} (b), we show an observable depending on two time-bin modes as a function of the time difference between the time-bin modes and for $N=20$. We can see how as $\Delta t$ diminishes the results computed with the Kraus map  (color lines) converge to those predicted by Eq. (\ref{analytical_two_ancilla}) which depend on two-time correlations that are computed with the master equation. In both panels we have considered the dynamics in the oscillatory regime and for long times, $n_1\gg1$, such that the system has already reached its stationary state.

\section{Mach-Zehnder interferometry of time-bin modes}\label{appendix_interferometer}

{\it Description of the interferometer. --} In order to describe the Mach-Zehnder interferometer of Fig. \ref{fig_cartoon} (b), we adopt the Heisenberg picture for the light field operators. We consider $50:50$ beam splitters with a $\pi/2$ phase difference between reflection and transmission. Then, the light field at the output of the first beam splitter is:
\begin{equation}
\begin{split}
\hat{a}_2(t)&=\frac{1}{\sqrt{2}}\big[\hat{a}_0(t)+i\hat{a}_1(t) \big],\\
\hat{a}_3(t)&=\frac{1}{\sqrt{2}}\big[i\hat{a}_0(t)+\hat{a}_1(t) \big],
\end{split}    
\end{equation}
where the subindex labels the different arms depicted in Fig. \ref{fig_cartoon} (b). After this, the fields travel through the different arms acquiring a different phase and time delay. Then, the output of the second beam splitter is given by:
\begin{equation}
\begin{split}
\hat{a}_4(t)&=\frac{1}{\sqrt{2}}\big[\hat{a}_2(t-\tau_2)e^{i\phi_2}+i\hat{a}_1(t-\tau_1)e^{i\phi_1} \big],\\
\hat{a}_5(t)&=\frac{1}{\sqrt{2}}\big[i\hat{a}_2(t-\tau_2)e^{i\phi_2}+\hat{a}_1(t-\tau_1)e^{i\phi_1}],
\end{split}    
\end{equation}
where the time delays $\tau_{1,2}$ are given by the optical length of each arm, and the phases correspond $\phi_{1,2}=\omega_0 \tau_{1,2}$. Notice that the relative phases pick up the carrier optical frequency $\omega_0$ around which  all our time scales are defined. Recall that an assumption of the input-output formalism is that $\omega_0\gg\omega,\Gamma$, by orders of magnitude. This difference in orders of magnitude makes the rapidly oscillating interference pattern associated to these phases not relevant for our problem. This is because our results concern timescales of the order of $\omega^{-1}$, which contain many cycles of $\omega_0$, hence allowing us to  freely select a point of the interference pattern around the desired time separation $\tau=\tau_1-\tau_2$. Nevertheless, for clarity, we keep accounting for these phases until the end of the derivation.

Relating the output fields to the input field, we obtain:
\begin{equation}
\begin{split}
\hat{a}_4(t)&=\frac{1}{2}\big[\hat{a}_0(t-\tau_2)e^{i\phi_2}-\hat{a}_0(t-\tau_1)e^{i\phi_1} \big]\\
&+\frac{i}{2}\big[ \hat{a}_1(t-\tau_2)e^{i\phi_2}+\hat{a}_1(t-\tau_1)e^{i\phi_1}\big],\\
\hat{a}_5(t)&=\frac{i}{2}\big[\hat{a}_0(t-\tau_2)e^{i\phi_2}+\hat{a}_0(t-\tau_1)e^{i\phi_1} \big]\\
&-\frac{1}{2}\big[ \hat{a}_1(t-\tau_2)e^{i\phi_2}-\hat{a}_1(t-\tau_1)e^{i\phi_1}\big].
\end{split}    
\end{equation}
We now assume that $\tau_2<\tau_1$. We proceed with the discretization in time of the light field, making the correspondence $t-\tau_1=n_1\Delta t$, $t-\tau_2=n_2\Delta t$ with $n_2>n_1$. The input of arm '0' contains the field emitted by the system, while the input of arm '1' is the vacuum. Then, we arrive to the following expression for the output of the Mach-Zehnder interferometer written in terms of time-bin modes:
\begin{equation}
\begin{split}
\hat{a}_4(t)&=\frac{1}{2}\big[b_{[n_2]}e^{i\phi_2}-b_{[n_1]}e^{i\phi_1} \big]
+\frac{i}{2}\big[ c_{[n_2]}e^{i\phi_2}+c_{[n_1]}e^{i\phi_1}\big],\\
\hat{a}_5(t)&=\frac{i}{2}\big[b_{[n_2]}e^{i\phi_2}+b_{[n_1]}e^{i\phi_1} \big]
-\frac{1}{2}\big[ c_{[n_2]}e^{i\phi_2}-c_{[n_1]}e^{i\phi_1}\big],
\end{split}    
\end{equation}
where $t=n\Delta t$, with $n>n_1,n_2$, and $c_{[n_1]}$ and $c_{[n_2]}$ are the annihilation operators for vacuum time-bin modes at times $n_1\Delta t$  and $n_2\Delta t$, respectively. Therefore, the interferometer is probing the following state:
\begin{equation}
\hat{\mu}_{[n_2,n_1]}\otimes |0_{c,n_2}\rangle \langle   0_{c,n_2}| \otimes |0_{c,n_1}\rangle \langle   0_{c,n_1}|.
\end{equation}
As the vacuum modes 'c' are independent of the parameter we want to estimate, the QFI of this state is the same as that of $\hat{\mu}_{[n_2,n_1]}$. This is because the QFI of a product between states is the sum of QFIs of each state of the product  \cite{Braun2018}, while the QFI for the vacuum state of modes 'c' is zero.

\begin{figure}[t!]
 \centering
 \includegraphics[width=1\columnwidth]{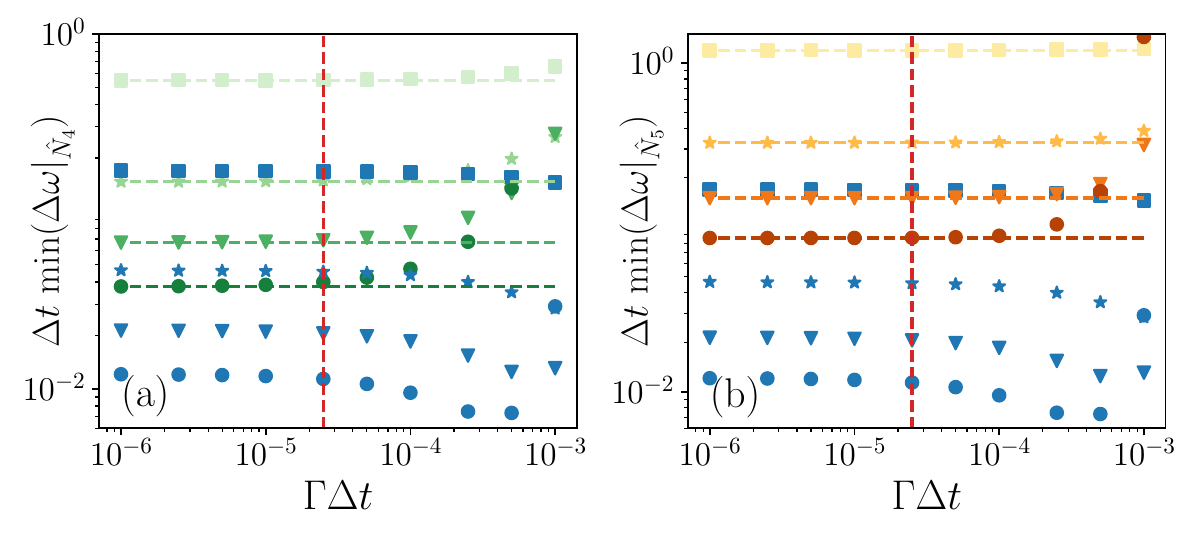}
 \caption{{\bf Counting measurements in the small-$\Delta t$ regime. }  (a) Green symbols: estimation error $\Delta\omega|_{\hat{N}_\mathrm{4}}$ at the optimal time $\tau^*$ varying $\Gamma \Delta t $ and $N$ ($N=20$ squares, $N=40$ stars, $N=60$ triangles, $N=80$ circles). Dashed green lines correspond to the approximate results of Eq. (\ref{equation_error_analytic}), obtained using the master equation two-time correlation functions. Blue symbols correspond to the fundamental bound for the corresponding parameters. (b) Orange symbols:  $\Delta\omega|_{\hat{N}_\mathrm{5}}$  for the same cases as in (a). In both panels $\omega/\omega_\mathrm{c}=2$.}
 \label{fig_Measurements_convergence}
\end{figure}

{\it Photon counting at the output arms. --} The photon counters depicted at the output arms of the interferometer, see Fig. \ref{fig_cartoon} (b), give us access to the following observables:
\begin{equation}
\begin{split}
\langle \hat{a}_4^\dagger  \hat{a}_4\rangle=&\frac{1}{4}\big[ \langle b^\dagger_{[n_1]}b_{[n_1]}\rangle + \langle b^\dagger_{[n_2]}b_{[n_2]}\rangle\\
&-\langle b^\dagger_{[n_1]}b_{[n_2]}\rangle e^{-i\Delta\phi}-\langle b^\dagger_{[n_2]}b_{[n_1]}\rangle e^{i\Delta\phi} \big] ,\\
\langle \hat{a}_5^\dagger  \hat{a}_5\rangle=&\frac{1}{4}\big[ \langle b^\dagger_{[n_1]}b_{[n_1]}\rangle + \langle b^\dagger_{[n_2]}b_{[n_2]}\rangle\\
&+\langle b^\dagger_{[n_1]}b_{[n_2]}\rangle e^{-i\Delta\phi}+\langle b^\dagger_{[n_2]}b_{[n_1]}\rangle e^{i\Delta\phi} \big],
\end{split}
\end{equation}
where expectation values $\langle \dots\rangle$ are taken with respect to the two time-bin mode reduced state, and we have dropped the irrelevant label $t$. Notice that the vacuum modes 'c' do not contribute to normal ordered observables. The substraction of the signal of both counters gives us access to the following observable:
\begin{equation}
\hat{N}_\mathrm{d}=\hat{a}_5^\dagger \hat{a}_5-  \hat{a}_4^\dagger \hat{a}_4, 
\end{equation}
whose expectation value has been presented in the main text. In order to compute the estimation error, we need to compute the following expectation values:
\begin{equation}
\begin{split}
\langle (\hat{a}_4^\dagger \hat{a}_4)^2\rangle&=   \langle \hat{a}_4^\dagger\hat{a}_4^\dagger \hat{a}_4\hat{a}_4\rangle+ \langle \hat{a}_4^\dagger\hat{a}_4\rangle,\\
\langle (\hat{a}_5^\dagger \hat{a}_5)^2\rangle&=   \langle \hat{a}_5^\dagger\hat{a}_5^\dagger \hat{a}_5\hat{a}_5\rangle+ \langle \hat{a}_5^\dagger\hat{a}_5\rangle,
\end{split}    
\end{equation}
which we have conveniently rewritten in normal order such that the vacuum modes 'c' do not contribute. Finally, in the small-$\Delta t$ regime, we can approximate the reduced two time-bin state by Eq. (\ref{analytical_two_ancilla}), which neglects two-photon transitions. From the dominant terms of Eq. (\ref{analytical_two_ancilla}), we observe that in this limit  $\langle (\hat{a}_4^\dagger \hat{a}_4)^2\rangle\approx   \langle \hat{a}_4^\dagger\hat{a}_4\rangle$ and $\langle (\hat{a}_5^\dagger \hat{a}_5)^2\rangle\approx   \langle \hat{a}_5^\dagger\hat{a}_5\rangle$. This can be used to obtain the approximate formulas for the estimation error given in the main text. In Fig. \ref{fig_measurements}, the approximated formula for the estimation error of associated with $\hat{N}_\mathrm{d}$ is benchmarked. In Fig. \ref{fig_Measurements_convergence}, we benchmark the expressions for the estimation errors of $\hat{N}_\mathrm{4,5}$. Similarly to the case of $\hat{N}_\mathrm{d}$, we observe that the results of numerically integrating the exact dynamics (color points) converge to the results  given by Eq. (\ref{equation_error_analytic}) for $N^2\Gamma\Delta t \ll 1$.

\section{Ansatz for the two-time correlations in the oscillatory regime}\label{appendix_ansatz}

\begin{figure}[t!]
 \centering
 \includegraphics[width=1\columnwidth]{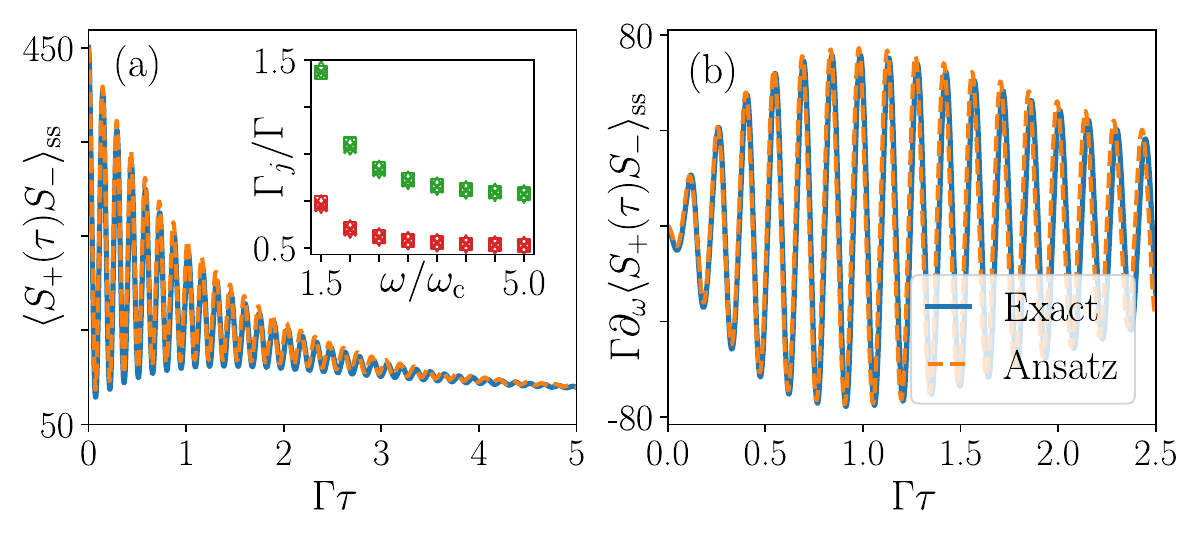}
 \caption{{\bf Ansatz for the two-time correlations in the oscillatory regime.} (a) Two time correlation $\langle S_+(\tau)S_-\rangle_\mathrm{ss}$ computed using the exact dynamics (blue solid line) or the ansatz of Eq. (\ref{ansatz_two_time}) (orange dashed line).  (b) Derivative of the two time correlation using the exact dynamics (blue solid line) or Eq. (\ref{ansatz_derivative}) (orange dashed line). Parameters $N=50$, $\omega/\omega_\mathrm{c}=2$. Inset: the two smallest (nonzero) decay rates of the Liouvillian, $\Gamma_1$ (red) and $\Gamma_2$ (green) for $N=20$ (squares), $N=50$ (diamonds) and $N=100$ (crosses). The values are very similar for different system sizes.}
 \label{fig_ansatz}
\end{figure}

In this section we provide a simple ansatz for the two-time correlation $\langle S_+(\tau)S_-\rangle_\mathrm{ss}$ in the oscillatory regime. We can write this two-time correlation in terms of the Liouvillian eigenmodes:
\begin{equation}
\langle S_+(\tau)S_-\rangle_\mathrm{ss}=|\langle S_+\rangle_\mathrm{ss}|^2+\sum_{j\geq 1}c_j e^{\lambda_j \tau}.    
\end{equation}
The coefficients $c_j$ and eigenvalues $\lambda_j$ can be obtained numerically. Our ansatz is based on keeping only the three first terms of the sum, i.e. those with the smallest decay rate, and propose a guess for their value. The eigenvalues are given by $\lambda_1=-\Gamma_1$, $\lambda_{2,3}\approx \pm i \Omega-\Gamma_2$, where $\Omega=\sqrt{\omega^2-\omega_\mathrm{c}^2}$ is the frequency of the mean-field oscillatory solution. We numerically find $\Gamma_{1,2}$ not to vary significantly with system size and to display values of the order of $\Gamma$, see inset of Fig. \ref{fig_ansatz} (a). The next step is to make a guess for the coefficients $c_{1,2,3}$. Based on the fact that at $\tau=0$ the two-time correlation takes the value $\langle S_+S_-\rangle_\mathrm{ss}$, we propose: $c_1=I_\mathrm{inc}/(2\Gamma)$, $c_2=I_\mathrm{inc}/(4\Gamma)$ and $c_3=c_2^*$, with $I_\mathrm{inc}=\Gamma(\langle S_+S_-\rangle_\mathrm{ss}-\langle S_+\rangle_\mathrm{ss}\langle S_-\rangle_\mathrm{ss})$. We compute the involved stationary expectation values numerically, although one could use the analytical expressions of Ref. \cite{Carmichael1980} which are already accurate for the considered sizes (not shown here). All together, our ansatz reads:
\begin{equation}\label{ansatz_two_time}
\langle S_+(\tau)S_-\rangle_\mathrm{ss}\approx|\langle S_+\rangle_\mathrm{ss}|^2 +\frac{I_\mathrm{inc}}{2\Gamma}( e^{-\Gamma_1\tau}+\cos(\Omega\tau)e^{-\Gamma_2\tau}).  
\end{equation}
We remark that with this ansatz we do not aim for perfect quantitative agreement but rather for an insightful formula that captures the main features of the collective response. This expression is benchmarked in Fig. \ref{fig_ansatz} (a) for the case $N=50$ and $\omega/\omega_\mathrm{c}=2$ finding good qualitative and quantitative agreement. We find a similar level of agreement when considering different system sizes (not shown here). We conclude that Eq. (\ref{ansatz_two_time}) captures the main features of this correlation  dynamics.

We now analyze the derivative with respect to $\omega$ of Eq. (\ref{ansatz_two_time}). In principle, we should consider the derivative of all  stationary expectation values as well as $\Gamma_{1,2}$. However, away from the phase transition, their contribution is small.  In the spirit of keeping only the dominant terms we neglect these terms, obtaining:
\begin{equation}
\partial_\omega\langle S_+(\tau)S_-\rangle_\mathrm{ss}\approx-\frac{I_\mathrm{inc}\tau \partial_\omega \Omega}{2\Gamma}\sin(\Omega\tau)e^{-\Gamma_2\tau},
\end{equation}
which is the expression given in the main text. As we can observe in Fig. \ref{fig_ansatz} (b), this approximate expression captures the main features of the exact dynamics, although it becomes less accurate for large $\tau$. A similar level of agreement is found for different system sizes. 

\section{Additional results with local decay}\label{Appendix_local_decay}

\begin{figure}[t!]
 \centering
 \includegraphics[width=1\columnwidth]{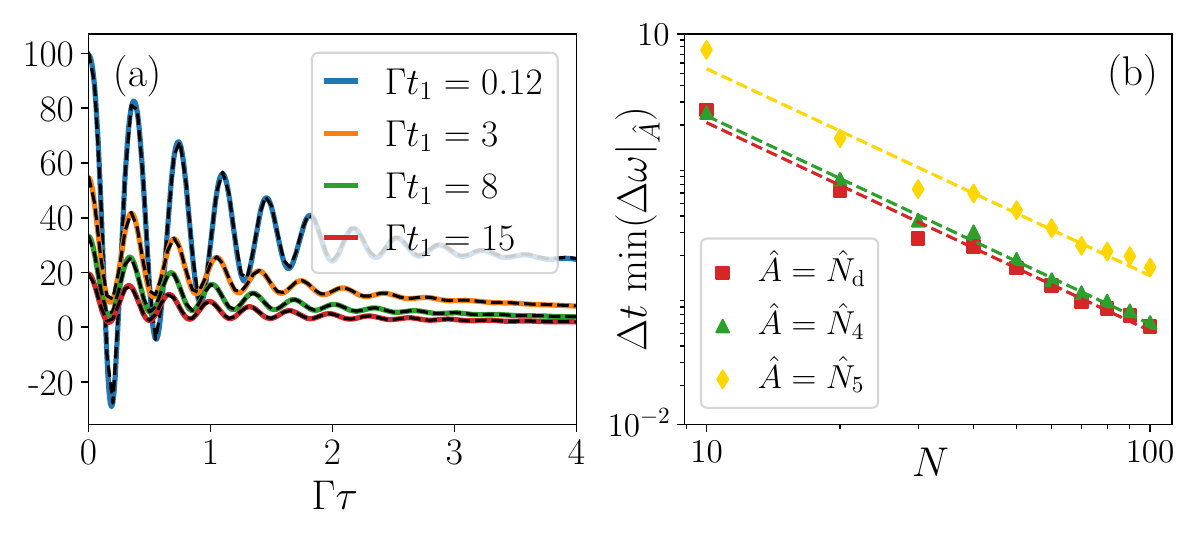}
 \caption{{\bf Additional results in the presence of local decay} (a) Two time-bin mode observables for the small-$\Delta t$ regime and local decay for  $\omega/\omega_\mathrm{c}=2$, $N=20$, $\gamma/\Gamma=0.1$, $\Gamma\Delta t=10^{-4}$ and initial condition all atoms in the ground state. Color solid lines correspond to the results integrating the discrete time dynamics. Black dashed lines correspond to the approximate results obtained from Eq. (\ref{analytical_two_ancilla}) computing the expectation values and two-time correlations with the master equation including local decay. (b) Minimum estimation errors with respect to $\tau$, and taking $\Gamma t_1=3$. The results of this panel are obtained from Eq. (\ref{equation_error_analytic_local_decay}). The dashed lines correspond to fits of the type $\propto N^{-\alpha}$ with $\alpha=(1.60,1.60,1.58)$ for $\hat{N}_\mathrm{d}$, $\hat{N}_4$ and $\hat{N}_5$, respectively. In all cases we take as  initial condition all atoms in the ground state.}
 \label{fig_local_decay_appendix}
\end{figure}

{\it small-$\Delta t$ regime. --} The approximate expression for the two time-bin mode reduced state given in Eq. (\ref{analytical_two_ancilla}) can also be used to analyze the small-$\Delta t$ regime in the presence of local dissipation. In such a case, one just needs to evaluate the expectation values and two-time correlations using the master equation with local decay, i.e. $\langle O\rangle_{t}=\text{Tr}\{O e^{(\mathcal{L}+\mathcal{L}_\mathrm{loc})t} \rho(0) \}$ and $\langle O_1(\tau)O_2\rangle_{t_1}=\text{Tr}\{O_1 e^{(\mathcal{L}+\mathcal{L}_\mathrm{loc})\tau} \big[O_2\rho(t_1) \big] \}$. In Fig. \ref{fig_local_decay_appendix} (a) we compare the two-time bin correlation $\langle b^\dagger_{[n_2]}b_{[n_1]}\rangle$ integrating the discrete-time dynamics (color lines) with the corresponding results obtained from adapting Eq. (\ref{analytical_two_ancilla}) to the presence of local losses (black-dashed lines). We show the case of $\omega/\omega_\mathrm{c}=2$, observing good agreement. Regarding the observed dynamics, we find that the smaller is $\Gamma t_1$ the larger is the two time-bin correlation. Moreover, we observe that around $\Gamma \tau\sim 3$ there is a change in the behavior of $\langle b^\dagger_{[n_2]}b_{[n_1]}\rangle$, which comes from the effects of local decay on the dynamics.

{\it Estimation error formulas in the small-$\Delta t$ regime. --} We can generalize the results of Eq. (\ref{equation_error_analytic}) to the case in which we have local decay and for finite $t_1=n_1\Delta t$. In order to do so, we proceed as before, by computing the two time-bin reduced state of Eq. (\ref{analytical_two_ancilla}) with the master equation containing local loses. The expressions for the estimation errors of the counting observables are given by:  
\begin{equation}\label{equation_error_analytic_local_decay}
\begin{split}
\Delta \omega|_{\hat{N}_\mathrm{d}}&\approx
\frac{1}{2\Gamma\Delta t}\frac{\sum_{j=1,2}\langle S_+ S_-\rangle_\mathrm{t_j}}{|\partial_\omega\text{Re}[\langle S_+(\tau)S_-\rangle_{t_1}]|^2},\\
\Delta \omega|_{\hat{N}_\mathrm{4,5}}&\approx
\frac{2}{\Gamma\Delta t}\frac{\frac{1}{2}\sum_{j=1,2}\langle S_+ S_-\rangle_\mathrm{t_j}\mp\text{Re}[\langle S_+(\tau)S_-\rangle_{t_1}]}{|\partial_\omega(\frac{1}{2}\sum_{j=1,2}\langle S_+ S_-\rangle_\mathrm{t_j}\mp\text{Re}[\langle S_+(\tau)S_-\rangle_{t_1}])|^2}.
\end{split}
\end{equation}
As discussed in the main text, the results are now sensitive to the choice of $\Gamma t_1$. In particular, in Fig. \ref{fig_local_decay_appendix} (b) we show results for $\Gamma t_1=3$. We observe that the scaling $N^{-2}$ is lost. Instead, one observes the approximate scaling law $N^{-1.6}$ which still surpasses the standard quantum limit. The results presented in Fig. \ref{fig_local_decay_appendix} (b) display some non-monotonous behavior with $N$. This is because when fixing $\Gamma t_1=3$ and varying $N$, the point of the sinusoidal pattern of $\langle S_\mathrm{y}\rangle_{t_1}$ that we are resolving varies with $N$.

\bibliography{references}

\end{document}